\documentclass[12pt]{article}

\usepackage[a4paper,left=1in,right=1in,top=1in,bottom=1in]{geometry}

\usepackage{amsmath,amssymb}
\usepackage{mathtools}
\usepackage{bm}

\usepackage{graphicx}
\usepackage{booktabs}

\usepackage{algorithm}
\usepackage{algorithmic}

\usepackage[numbers,sort&compress]{natbib}
\usepackage[hidelinks]{hyperref}

\title{Unsteady airfoil aerodynamics in attached flow: From unsteady thin airfoil theory to wind turbine application}

\author{%
  Ang Li\thanks{Corresponding author: \texttt{angl@dtu.dk}}
  \qquad Mac Gaunaa
  \qquad Georg Raimund Pirrung\\[0.6ex]
  \small Department of Wind and Energy Systems, Technical University of Denmark\\
  \small Frederiksborgvej 399, 4000 Roskilde, Denmark
}
\date{}

\begin{document}

\maketitle

\begin{abstract}
Attached-flow unsteady aerodynamics form the foundation of the dynamic stall models used in wind turbine aerodynamic and aeroelastic codes.
Such conditions are particularly important over the outboard blade region, which dominates power production and aerodynamic loading.
In this region, the angle of attack and relative velocity are unsteady while the flow remains predominantly attached.
A complete engineering model for unsteady airfoil aerodynamic loads combines several distinct elements, including airfoil polar lookup, attached-flow shed wake memory, non-circulatory loads, consistent definitions of force magnitude and direction, and, subsequently, separated-flow dynamics.
Although the underlying theory is classical, to the authors' knowledge, existing descriptions do not provide a complete and internally consistent route for implementing these attached-flow contributions in wind turbine aerodynamic solvers.
This work focuses on the attached-flow component and presents all attached-flow contributions in the form of lift, drag, and moment coefficients for implementation in generalized lifting-line methods, including blade-element momentum (BEM), lifting-line (LL), and actuator-line (AL) methods.
The formulation is designed for combination with 2-D airfoil polars obtained from measurements or CFD.
Starting from the classical dimensional loads of unsteady thin airfoil theory, the circulatory and non-circulatory contributions are derived in coefficient form, and the implications of the associated modeling choices are clarified.
In particular, the shed wake memory is formulated using the downwash velocity rather than angle of attack as the aerodynamic state variable.
Because the 2-D theory is classical, verification focuses on its rotor-level implementation.
The coned straight blade case provides a cross-method benchmark and quantifies the errors in rotor-integrated thrust and power caused by omitting three required but often omitted contributions.
The lift direction projection contribution has the largest influence on power, while omitting the mid-chord heaving-acceleration contribution eliminates the non-circulatory normal-force cancellation and produces a thrust error.
The zero-onset flow vertical-axis wind turbine (VAWT) case verifies that, in the ideal thin-airfoil limit, all circulatory and non-circulatory contributions cancel, resulting in zero total rotor torque.
\end{abstract}

\section{Introduction}
\label{sec:intro}

For modern wind turbines, both the relative velocity and angle of attack at a blade section can be unsteady even when the local flow remains attached.
Such conditions are common, especially for the outboard blade region that dominates power production and aerodynamic loading for horizontal-axis wind turbines (HAWTs).
The effects described by the attached-flow part of dynamic stall models are particularly relevant for large rotors with highly coned or deflected blades, where sectional unsteady aerodynamic modeling can be important even under steady operating conditions~\citep{Li2022_How}.
In vertical-axis wind turbines (VAWTs), both the relative velocity and the angle of attack vary continuously with the azimuth position of the blades.
Dynamic stall models in wind turbine aeroelastic codes contain several distinct modeling elements: airfoil polar lookup, shed wake memory in attached flow, non-circulatory loads, the definition of force magnitude and direction, and finally separated flow dynamics that are typically built on the attached flow formulation.
The present work focuses on the attached-flow component and formulates the circulatory and non-circulatory loads as force and moment coefficients that can be combined with 2-D airfoil polars from measurements or CFD simulations.
When the formulation is embedded in a dynamic-stall model, the attached-flow formulation provides the basis on which the separated-flow dynamics are subsequently introduced, while the separated-flow aerodynamic loads themselves remain outside the scope of the present work.
The airfoil polar provides the quasi-steady lift magnitude, profile drag, and pitching moment at the effective angle of attack.
The attached-flow model accounts for the effect of the shed wake memory and specifies how the resulting circulatory force is resolved into lift and drag relative to a local reference flow direction.
Different reference directions change the lift and drag decomposition, but not the underlying aerodynamic force vector when the projection is treated consistently.
The non-circulatory terms of the model provide additional force and moment contributions associated with the inertia of the fluid accelerated by the airfoil motion.
These contributions are commonly referred to as added-mass or apparent-mass effects and arise naturally as the non-circulatory part of classical unsteady thin-airfoil theory.
They may equivalently be interpreted as the potential-flow reaction forces generated by accelerating the fluid surrounding a body.

Although the underlying theory is classical and individual elements of the formulation have appeared in previous studies, to the authors' knowledge, the wind turbine aerodynamics literature still lacks a complete, internally consistent, and implementation-ready description.
The present work addresses this documentation gap by presenting a complete coefficient formulation that specifies how the circulatory lag state, airfoil polar lookup, force projection, and non-circulatory terms are combined consistently and without double counting.

The circulatory force magnitude and the memory effect of the shed wake follow classical unsteady thin airfoil theory, as represented by Wagner's indicial response and Theodorsen's frequency-domain formulation~\citep{Theodorsen1935,Jones1938,Jones1940}.
The tangential force contribution required to determine the resulting force direction follows the classical leading edge suction treatment of \citet{Garrick1936} and its generalized thin-airfoil formulation by \citet{Gaunaa2010}.
The non-circulatory loads also follow from unsteady thin airfoil potential flow theory, including its extension to time-varying inflow velocity~\citep{Garrick1936,Greenberg1947}.
The implementations discussed here are based on the model documented by \citet{Hansen2004}, commonly known as the Risø model and termed the Hansen-Gaunaa-Madsen (HGM) model in \citet{BranlardJonkman2019}, a name subsequently used in the OpenFAST literature~\citep{Branlard2022}.
For the low Mach numbers relevant to wind turbine applications, \citet{Hansen2004} specialised the Beddoes-Leishman dynamic stall model~\citep{LeishmanBeddoes1986,Leishman1989_semi} to incompressible aerodynamics by replacing the compressible impulsive response with non-circulatory added-mass loads.
The HGM model also excludes leading edge separation and the associated leading edge vortex dynamics.
In the original formulation, some terms of the underlying attached-flow theory were deliberately not retained in the engineering model, as they were considered of minor importance for the operating conditions and predominantly planar and relatively stiff HAWT rotors targeted at the time.
Indeed, \citet{Hansen2004} derived the complete non-circulatory formulation before neglecting several contributions, such as the mid-chord heaving acceleration, in the reduced engineering model.
As the model has subsequently been applied to more flexible and increasingly non-planar rotors, including highly coned or deflected configurations~\citep{Li2022_How}, and to other concepts such as VAWTs~\citep{Pirrung2018}, the consequences of these simplifications have become more apparent.
This provides part of the motivation for the present work, which aims to provide the complete attached-flow framework together with a detailed specification of its inputs, force and moment definitions, reference directions, and implementation.

Subsequent work has further developed this incompressible Beddoes-Leishman type dynamic stall model for HAWC2~\citep{Bergami2012ATEFlap,Pirrung2018}; related formulations are used in several widely used wind-turbine aeroelastic simulation tools, including OpenFAST \texttt{UAMod=4}~\citep{Branlard2022}, the incompressible Beddoes-Leishman model in Bladed~\citep{DNVBladed2025}, and the ATEFlap model~\citep{Bergami2012ATEFlap} as implemented in QBlade~\citep{Marten2020QBlade}.
Related formulations of the incompressible attached-flow unsteady aerodynamics have also been applied in typical section aeroelastic models~\citep{Stablein2017,Verdonck2026}.
Although these implementations are based on the same underlying model family, their descriptions differ in the definition of the circulatory lag state, the non-circulatory terms retained, and the conventions used to express the resulting force and moment coefficients.

Two aspects of the implementation require particular attention.
The first is the state variable used for the circulatory lag.
The present formulation expresses the circulatory lag using downwash-based states rather than angle-of-attack states.
A deficit-state representation is used that, when embedded in the full dynamic-stall model, allows the generation of new wake memory to be progressively reduced as flow separation develops.
The second aspect is the complete set of non-circulatory force and moment terms required when the formulation is written directly in terms of lift, drag, and quarter-chord pitching moment coefficients.
These terms must also be transformed consistently into the selected lift and drag directions.

The model is applicable in all generalized lifting-line methods: aerodynamic models that discretize the blade into sections and use 2-D airfoil data, including blade-element momentum (BEM), lifting-line (LL), and actuator-line (AL) methods.
These methods combine an outer 3-D rotor and wake problem with an inner 2-D airfoil model at each blade section.
In simplified implementations, the inner 2-D model is often reduced to a direct airfoil-polar lookup using the flow at a single aerodynamic calculation point. Careful analysis of the thin airfoil framework, however, shows that the flow angle at a single point is not, by itself, sufficient to predict the unsteady attached flow forces correctly~\citep{Bergami2012ATEFlap,Pirrung2018,Li2022_How}. The local flow angles at both the quarter- and three-quarter-chord points are required. These angles are generally different because the effective torsional rate causes the local flow angle to vary along the chord. They must therefore be determined from the flow at the aerodynamic calculation point together with the effective torsional rate.
For horizontal-axis wind turbines (HAWTs) with out-of-plane blade geometry and for VAWTs, inconsistent treatment of the non-circulatory loads or neglect of the effective torsional rate can produce erroneous loads~\citep{Pirrung2018,Li2022_How}.
The classical 2-D sectional responses underlying the formulation have been extensively studied in previous work. Verification therefore focuses on its implementation in rotor aerodynamic solvers, where the blade kinematics must be consistently coupled to the sectional aerodynamic model and the resulting loads transformed back to the rotor system.
The present work uses two complementary rotor-level cases.
The coned straight-blade case provides a cross-method numerical benchmark and quantifies the errors caused by omitting required force-projection and non-circulatory contributions.
The zero-onset VAWT case provides an analytical consistency test building on previous studies of the circulatory torque balance and non-circulatory normal-force cancellation~\citep{Pirrung2018,Li2022_How}.
The present work extends these analyses by considering all circulatory and non-circulatory contributions and demonstrates that the complete formulation results in zero total aerodynamic torque when using ideal zero profile-drag airfoil data.

The derivation starts from the classical sectional normal force, tangential force, and pitching moment and proceeds through chord-axis coefficients.
The final formulation gives lift and drag coefficients in the selected flow direction and the pitching moment coefficient about the quarter-chord point.
The remainder of the paper derives the coefficient equations in Sect.~\ref{sec:methodology}, examines the two rotor-level cases in Sect.~\ref{sec:verification}, provides the time-discrete downwash- and deficit-state updates in Appendix~\ref{app:indicial-lag-updates}, and concludes in Sect.~\ref{sec:conclusion}.

\section{Methodology: from classical sectional loads to engineering coefficients}
\label{sec:methodology}

This section derives a set of aerodynamic coefficient equations suitable for coupling the inner 2-D airfoil model to generalized lifting-line methods.
The first part presents the classical dimensional normal force, tangential force, and pitching moment about an arbitrary chordwise attachment point and expresses these loads in non-dimensional coefficient form. 
The pitching moment coefficient is subsequently re-referenced to the quarter-chord point, and the shed wake memory is then written in state-space form.
The second part develops the practical formulation required to combine these results with a 2-D airfoil polar.
It couples the outer blade and wake model to the inner 2-D airfoil model, interprets the circulatory force magnitude and direction, introduces the airfoil polar lookup, and projects the force contributions into lift and drag relative to a local reference flow direction.
The resulting contributions are then assembled into the total lift, drag, and pitching moment coefficients.

\subsection{Classical dimensional unsteady thin airfoil theory}
\label{sec:dimensional-thin-airfoil-loads}

The classical formulation considers incompressible 2-D potential flow under the small-angle assumptions of thin airfoil theory \cite{Theodorsen1935,Garrick1936,Oye1981,Gaunaa2010}.
Airfoil-specific viscous effects are introduced later through coupling with a steady 2-D airfoil polar.

\begin{figure}[!htbp]
  \centering
  \includegraphics[width=0.5\linewidth]{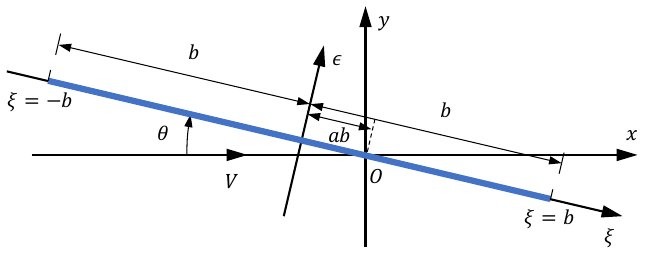}
  \caption{Definitions of the flow and airfoil coordinate systems and positive directions used in the derivation of unsteady 2-D thin airfoil theory. Adapted from \citet{Li2022_How} with permission.}
  \label{fig:airfoil-coordinate}
\end{figure}

The flow and airfoil coordinate systems and the positive directions used in the derivation are shown in Fig.~\ref{fig:airfoil-coordinate}.
The coordinate system and sign conventions follow \citet{Li2022_How}.
Throughout this work, $c$ denotes the chord length and $b=c/2$ the half-chord length.
The chordwise coordinate is denoted by $\xi$ and is measured from the mid-chord point, with $\xi=-b$ at the leading edge and $\xi=b$ at the trailing edge.
The pitch axis is located at $\xi=ab$, where $a$ is its dimensionless position relative to the mid-chord point.
The motion of the airfoil is described by the translational displacements $x$ and $y$, parallel and normal to the undisturbed flow direction, respectively, and by the pitching angle $\theta$, which describes the torsional motion.

The relative streamwise velocity is defined as
\begin{equation}
  U=V-\dot{x},
  \label{eq:relative-velocity}
\end{equation}
where $V$ is the constant streamwise flow velocity.

The classical literature on unsteady thin airfoil theory adopts the simplifying assumption that the influence of the shed wake at any chordwise position is evaluated as if the airfoil and the shed wake lie along a straight line, with the shed vorticity convected downstream relative to the airfoil at the streamwise velocity $U(t)$.
The theory is developed by enforcing the Neumann boundary condition, corresponding to no flow through the airfoil, together with the Kutta condition, which removes the pressure singularity at the trailing edge.
The unsteady Bernoulli equation is then used to evaluate the local pressure difference, which is integrated to obtain the sectional aerodynamic loads.
Because of the leading edge singularity in the thin airfoil theory, the associated leading edge suction force must be included to obtain the correct tangential load \citep{Garrick1936,Gaunaa2010}.
The suction force may be interpreted as the limiting physical suction force as the airfoil thickness tends to zero \citep{Gaunaa2010}. Its inclusion recovers zero drag in the steady inviscid limit, consistent with d'Alembert's paradox.

Within this framework, the input to the unsteady thin airfoil model is the relative streamwise velocity $U(t)$ together with the onset downwash distribution $w(\xi,t)$, which is the undisturbed relative onset velocity component normal to the airfoil chord. In this framework, at each instant in time, the onset downwash varies linearly along the chord, as shown explicitly in Eq.~(\ref{eq:downwash-chordwise}). The downwash distribution can therefore be characterized by its value at one chordwise location and its chordwise slope, which is set by the effective torsional rate $\dot{\theta}$.
For use in generalized lifting-line models, the unsteady 2-D airfoil formulation may therefore be interpreted as an input--output machine. It maps the time histories of the relative flow magnitude $U$ and the chordwise-linear onset downwash to the corresponding time histories of the integral sectional loads, expressed below as the normal force, tangential force, and pitching moment.


Differentiation of Eq.~(\ref{eq:relative-velocity}) gives
\begin{equation}
  \dot{U}=-\ddot{x}.
  \label{eq:relative-velocity-rate}
\end{equation}
The formulation assumes a positive relative streamwise speed, $U>0$.
Under the small-angle assumptions of thin airfoil theory, the mid-chord heaving acceleration normal to the airfoil is
\begin{equation}
  \ddot{\epsilon}
  =
  \ddot{x}\theta+\ddot{y}+ab\ddot{\theta}.
  \label{eq:epsilon}
\end{equation}

The onset downwash at the three-quarter-chord point is approximated as
\begin{equation}
  w_{3/4}
  =
  U\theta-\dot{y}
  +
  \left(\frac{1}{2}-a\right)b\dot{\theta}.
  \label{eq:w34}
\end{equation}

The individual circulatory and non-circulatory contributions are identified from the dimensional loads.
The normal force $N$ acts perpendicular to the chord and is positive from the pressure side towards the suction side.
The tangential force $T$ acts along the chord and is positive towards the leading edge.
The pitching moment $M_a$ is taken about the pitch axis at $\xi=ab$ and is positive in the nose-up direction.

For harmonic motion at constant $U$, the effect of the shed wake is represented by Theodorsen's function $C(k)$.
Writing a harmonic downwash quantity as $w(t)=\Re\{\hat{w}e^{i\omega t}\}$, where the hat denotes the complex harmonic amplitude, the frequency-domain relation between the equivalent wake-lagged downwash and the three-quarter-chord onset downwash is
\begin{equation}
    \hat{w}_E
    =
    C(k)\hat{w}_{3/4},
    \label{eq:wE-frequency}
\end{equation}
where $k=\omega b/U$ is the reduced frequency.

In time domain simulations, $w_E$ is obtained from the indicial-response states introduced in Sect.~\ref{sec:wake-state}. It is an equivalent wake-lagged downwash used for determining the integrated circulatory loading in the model and should not be interpreted as a real physical wake-induced velocity evaluated at a particular chordwise position.

For the uncambered airfoil considered here, using $w_E$ and the mid-chord heaving acceleration defined in Eq.~(\ref{eq:epsilon}), the dimensional unsteady thin airfoil loads per unit span obtained from the general formulation given in Appendix B of \citet{Gaunaa2010} can be written in the present notation as
\begin{gather}
  N
  =
  \underbrace{
    2\pi\rho b\,U w_E
  }_{N_{\mathrm{circ}}}
  +
  \underbrace{
    \pi\rho b^2
    \left(
      U\dot{\theta}
      -
      \ddot{\epsilon}
    \right)
  }_{N_{\mathrm{NC}}},
  \label{eq:N-classical}
  \\
\begin{aligned}
  T
  &=
  2\pi\rho b
  \left(
    w_E-\frac{b}{2}\dot{\theta}
  \right)^2
  \\
  &=
  \underbrace{
    2\pi\rho b\,w_E^2
    -
    2\pi\rho b^2w_E\dot{\theta}
  }_{T_{\mathrm{circ}}^{\mathrm{suction}}}
  +
  \underbrace{
    \frac{1}{2}\pi\rho b^3\dot{\theta}^2
  }_{T_{\mathrm{NC}}^{\dot{\theta}^2}},
  \end{aligned}
  \label{eq:T-classical}
  \\
  M_a
  =
  \underbrace{
    2\pi\rho b^2
    \left(a+\frac{1}{2}\right)
    U w_E
  }_{M_{a,\mathrm{circ}}}
  +
  \underbrace{
    \pi\rho b^3a
    \left(
      U\dot{\theta}
      -
      \ddot{\epsilon}
    \right)
    -
    \frac{1}{2}\pi\rho b^3U\dot{\theta}
    -
    \frac{1}{8}\pi\rho b^4\ddot{\theta}
  }_{M_{a,\mathrm{NC}}}.
  \label{eq:Ma-classical}
\end{gather}

The non-circulatory normal force consists of the mid-chord heaving-acceleration and torsional-rate contributions
\begin{gather}
  N_{\mathrm{NC}}^{\mathrm{acc}}
  =
  -\pi\rho b^2\ddot{\epsilon},
  \label{eq:Nnc-acc-dimensional}\\
  N_{\mathrm{NC}}^{\mathrm{tor}}
  =
  \pi\rho b^2U\dot{\theta}.
  \label{eq:Nnc-tor-dimensional}
\end{gather}

For airfoil polars, the pitching moment is generally provided about the quarter-chord point.
The reference for the thin airfoil pitching moment is therefore transferred from the pitch axis at $\xi=ab$ to the quarter-chord point at $\xi=-b/2$:
\begin{equation}
  M_{1/4}^{\mathrm{thin}}
  =
  M_a
  -
  \left(a+\frac{1}{2}\right)bN.
  \label{eq:moment-transfer-quarter}
\end{equation}
Substitution of the normal force and pitching moment cancels the circulatory contribution exactly and gives
\begin{equation}
  M_{1/4}^{\mathrm{thin}}
  =
  -\frac{c}{4}N_{\mathrm{NC}}^{\mathrm{acc}}
  -
  \frac{c}{2}N_{\mathrm{NC}}^{\mathrm{tor}}
  -
  \frac{1}{8}\pi\rho b^4\ddot{\theta}.
  \label{eq:M-quarter-thin}
\end{equation}
Thus, for the uncambered thin airfoil considered here, the quarter-chord pitching moment contains only non-circulatory contributions.


\subsubsection{Physical interpretation of the non-circulatory terms}
\label{sec:physical-interpretation-nc}

The physical meaning of the non-circulatory terms can be clarified by considering the chordwise distribution of the onset downwash.
For an arbitrary chordwise position $\xi$, the small-angle rigid-airfoil kinematics give
\begin{equation}
  w(\xi,t)
  =
  U\theta-\dot{y}
  +
  (\xi-ab)\dot{\theta}
  =
  w_{1/2}(t)+\xi\dot{\theta}(t),
  \label{eq:downwash-chordwise}
\end{equation}
where $w_{1/2}=U\theta-\dot{y}-ab\dot{\theta}$ is the onset downwash at the mid-chord point.
Equation~(\ref{eq:downwash-chordwise}) shows that $\dot{\theta}$ determines the linear variation of the downwash along the chord.

Differentiating Eq.~(\ref{eq:downwash-chordwise}) gives
\begin{equation}
  \dot{w}(\xi,t)
  =
  \dot{w}_{1/2}(t)
  +
  \xi\ddot{\theta}(t).
  \label{eq:downwash-rate-chordwise}
\end{equation}
Using $\dot{U}=-\ddot{x}$ in Eq.~(\ref{eq:relative-velocity-rate}) together with Eq.~(\ref{eq:epsilon}), the rate of change of the mid-chord downwash is
\begin{equation}
  \dot{w}_{1/2}
  =
  U\dot{\theta}
  -
  \ddot{\epsilon}.
  \label{eq:downwash-rate-midchord}
\end{equation}

The combination of the two terms in the non-circulatory normal force can therefore be interpreted as the rate of change of the mid-chord downwash.
The term $-\ddot{\epsilon}$ represents the contribution from the mid-chord kinematic acceleration normal to the airfoil, whereas $U\dot{\theta}$ results from the changing projection of the streamwise relative flow as the airfoil rotates.
Although $U\dot{\theta}$ is uniform along the chord in $\dot{w}$, the same torsional rate $\dot{\theta}$ simultaneously determines the linear chordwise variation of $w$ through Eq.~(\ref{eq:downwash-chordwise}).

The torsional-rate-squared contribution $T_{\mathrm{NC}}^{\dot{\theta}^2}$ is a non-circulatory tangential load associated with the linear variation of downwash along the chord produced by $\dot{\theta}$.

The physical interpretation of the non-circulatory quarter-chord moment can be made more transparent by rewriting Eq.~(\ref{eq:M-quarter-thin}) as
\begin{equation}
  M_{1/4,\mathrm{NC}}
  =
  -\frac{c}{4}N_{\mathrm{NC}}
  -
  \frac{c}{4}N_{\mathrm{NC}}^{\mathrm{tor}}
  -
  \frac{1}{8}\pi\rho b^4\ddot{\theta}.
  \label{eq:M-quarter-nc-physical}
\end{equation}
The first term corresponds to placing the total non-circulatory normal force at the mid-chord point.
The second term is an additional moment associated with the linear variation of downwash along the chord caused by the torsional rate.
The final term is associated with the torsional acceleration $\ddot{\theta}$. For pure rotation about the mid-chord, $\ddot{\theta}$ produces opposite normal accelerations on the leading- and trailing-edge sides, giving no resultant normal force but a non-zero pitching moment.

\subsection{Thin airfoil loads in coefficient form}
\label{sec:thin-airfoil-coefficients}

The dynamic pressure and characteristic time for the flow to travel one half chord are
\begin{gather}
  q=\frac{1}{2}\rho U^2,
  \label{eq:dynamic-pressure}\\
  T_0=\frac{b}{U}=\frac{c}{2U}.
  \label{eq:characteristic-time}
\end{gather}

The equivalent angle of attack associated with the wake-lagged downwash is defined as
\begin{equation}
  \alpha_E
  =
  \frac{w_E}{U}.
  \label{eq:alphaE-definition}
\end{equation}
The equivalent angle of attack is a calculated quantity that governs the circulatory-force magnitude and should not be interpreted as an actual local flow angle at the three-quarter-chord point.

Non-dimensionalisation of Eq.~(\ref{eq:N-classical}) gives the normal force coefficient
\begin{equation}
  C_N
  =
  \frac{N}{qc}
  =
  2\pi\alpha_E
  +
  C_{N,\mathrm{NC}},
  \label{eq:CN-total}
\end{equation}
where the non-circulatory contribution is
\begin{equation}
  C_{N,\mathrm{NC}}
  =
  C_{N,\mathrm{NC}}^{\mathrm{acc}}
  +
  C_{N,\mathrm{NC}}^{\mathrm{tor}},
  \label{eq:cn}
\end{equation}
with
\begin{gather}
  C_{N,\mathrm{NC}}^{\mathrm{acc}}
  =
  -\pi T_0
  \frac{\ddot{\epsilon}}{U},
  \label{eq:cn-acc}
  \\
  C_{N,\mathrm{NC}}^{\mathrm{tor}}
  =
  \pi T_0\dot{\theta}.
  \label{eq:cn-tor}
\end{gather}

Non-dimensionalisation of Eq.~(\ref{eq:T-classical}) gives the tangential force coefficient
\begin{equation}
  C_T
  =
  \frac{T}{qc}
  =
  \underbrace{
    2\pi
    \alpha_E^2
    -
    2\pi T_0\dot{\theta}
    \alpha_E
  }_{
    C_{T,\mathrm{circ}}^{\mathrm{suction}}
  }
  +
  \underbrace{
    \frac{1}{2}\pi
    \left(
      T_0\dot{\theta}
    \right)^2
  }_{
    C_{T,\mathrm{NC}}^{\dot{\theta}^2}
  }.
  \label{eq:ct}
\end{equation}

The full non-circulatory pitching moment coefficient about the quarter-chord point is
\begin{equation}
  \begin{split}
    C_{m,\mathrm{NC}}^{\mathrm{full}}
    \equiv
    \frac{M_{1/4}^{\mathrm{thin}}}{qc^2}
    &=
    \frac{1}{4}\pi T_0
    \frac{\ddot{\epsilon}}{U}
    -
    \frac{1}{2}\pi T_0\dot{\theta}
    -
    \frac{1}{16}\pi T_0^2\ddot{\theta}
    \\
    &=
    -\frac{1}{4}
    C_{N,\mathrm{NC}}^{\mathrm{acc}}
    -
    \frac{1}{2}
    C_{N,\mathrm{NC}}^{\mathrm{tor}}
    -
    \frac{1}{16}\pi T_0^2\ddot{\theta}.
  \end{split}
  \label{eq:cm-full}
\end{equation}

\subsection{Shed wake memory in state-space form}
\label{sec:wake-state}

The frequency-domain shed wake relation in Eq.~(\ref{eq:wE-frequency}) is represented in the time domain through an exponential approximation of an indicial response.
For an approximation with $n_w$ exponential terms, the response function is written as
\begin{equation}
  \phi(s)
  =
  1-\sum_{i=1}^{n_w}A_i e^{-\beta_i s}.
  \label{eq:wagner-general}
\end{equation}
The non-dimensional convective time, which measures the distance travelled by the wake in half-chords, is defined as
\begin{equation}
  s
  =
  \frac{2}{c}
  \int_0^t
  U(t')\,\mathrm{d}t'.
  \label{eq:convective-time}
\end{equation}
Because $T_0=c/(2U)$, the rate of change of the convective time is $\mathrm{d}s/\mathrm{d}t=1/T_0$.

Two- and three-term approximations are commonly used in practice.
For example, the two-term approximation of Wagner's thin airfoil impulsive response by \citet{Jones1938,Jones1940} uses $n_w=2$, with $A_1=0.165$, $A_2=0.335$, $\beta_1=0.0455$, and $\beta_2=0.3$. More general impulsive response functions that account for effects of airfoil thickness and shape have also been developed \citep{Bergami2013}.

The following equations are first written in terms of the geometric three-quarter-chord downwash, consistent with the classical symmetric thin-airfoil formulation and the angle-state formulation of \citet{Hansen2004}.
Here,
\begin{gather}
  w_{3/4}
  =
  U\alpha_{3/4},
  \label{eq:geometric-downwash-definition}
  \\
  w_E
  =
  U\alpha_E.
  \label{eq:equivalent-downwash-definition}
\end{gather}
Here, $\alpha_E$ is the equivalent angle of attack. 
The treatment of cambered airfoils is introduced later through a zero-lift-shifted downwash.

In the direct downwash-state form, the equivalent wake-lagged downwash governing the circulatory response is written as
\begin{equation}
  w_E
  =
  \left(
    1-\sum_{i=1}^{n_w}A_i
  \right)
  w_{3/4}
  +
  \sum_{i=1}^{n_w}x_i^w.
  \label{eq:wE-states}
\end{equation}
Each state $x_i^w$ represents one component of the wake-induced memory and satisfies
\begin{equation}
  \dot{x}_i^w
  +
  \frac{\beta_i}{T_0}x_i^w
  =
  \frac{\beta_iA_i}{T_0}w_{3/4}.
  \label{eq:w-state-ode}
\end{equation}

Starting from the downwash-state formulation, \citet{Hansen2004} introduces an equivalent formulation in terms of angle-of-attack states.
The corresponding angle state $z_i$ is defined by
\begin{equation}
  x_i^w
  =
  Uz_i.
  \label{eq:angle-state-transform}
\end{equation}
Substitution into Eq.~(\ref{eq:wE-states}), followed by division by $U$ and use of Eqs.~(\ref{eq:geometric-downwash-definition}) and (\ref{eq:equivalent-downwash-definition}), gives
\begin{equation}
  \alpha_E
  =
  \left(
    1-\sum_{i=1}^{n_w}A_i
  \right)
  \alpha_{3/4}
  +
  \sum_{i=1}^{n_w}z_i.
  \label{eq:alphaE-angle-states}
\end{equation}
In this angle-state formulation, $\alpha_E$ is not shifted by the zero-lift angle $\alpha_0$.
Differentiating Eq.~(\ref{eq:angle-state-transform}) and substituting the result into Eq.~(\ref{eq:w-state-ode}) give
\begin{equation}
  \dot{z}_i
  +
  \left(
    \frac{\beta_i}{T_0}
    +
    \frac{\dot{U}}{U}
  \right)z_i
  =
  \frac{\beta_iA_i}{T_0}\alpha_{3/4}.
  \label{eq:angle-state-ode}
\end{equation}
For time-varying $U$, the angle-state formulation requires the additional term $\dot{U}/U$, whereas the downwash-state formulation does not.

The direct downwash-state and angle-state forms are mathematically equivalent descriptions of the attached-flow shed wake memory when the velocity-rate term and the initial conditions are treated consistently.
A further reformulation is useful when the attached-flow model is embedded in a dynamic-stall model because it separates the decay of stored wake memory from the new contribution to the attached-flow lag.

The preceding formulations use the three-quarter-chord angle of attack $\alpha_{3/4}$ and the corresponding onset downwash $w_{3/4}=U\alpha_{3/4}$.
For the classical symmetric thin airfoil, the zero-lift angle is $\alpha_0=0$.
For a cambered airfoil, $\alpha_0$ denotes the zero-lift angle of the 2-D airfoil polar.
The deficit-state formulation introduced by \citet{Pirrung2018} uses the three-quarter-chord onset downwash $w_{3/4}=U\alpha_{3/4}$ as input.
The airfoil zero-lift angle is not considered explicitly in that formulation.
Following \citet{Pirrung2017Comparison}, the zero-lift offset of a cambered airfoil is included in the onset-downwash input to the shed wake model. In the present downwash-state formulation, this is written as:
\begin{equation}
  \widetilde{w}_{3/4}
  =
  U\left(\alpha_{3/4}-\alpha_0\right)
  =
  w_{3/4}-U\alpha_0.
  \label{eq:shifted-w34}
\end{equation}
The tilde denotes an onset-downwash quantity measured relative to the zero-lift condition.
The consistency of this shift for time-varying $U$ is demonstrated in Appendix~\ref{sec:appendix-zero-lift-offset}.

When the direct downwash-state equations are formulated using $\widetilde{w}_{3/4}$, the deficit states are defined as
\begin{equation}
  r_i^w
  =
  A_i\widetilde{w}_{3/4}-x_i^w.
  \label{eq:w-deficit-transform}
\end{equation}
The corresponding zero-lift-shifted equivalent wake-lagged downwash is
\begin{equation}
  \widetilde{w}_E
  =
  \widetilde{w}_{3/4}
  -
  \sum_{i=1}^{n_w}r_i^w.
  \label{eq:wE-deficit-states}
\end{equation}
The corresponding effective angle of attack is then recovered as
\begin{equation}
  \alpha_E
  =
  \alpha_0
  +
  \frac{\widetilde{w}_E}{U}.
  \label{eq:alphaE-zero-lift-shift}
\end{equation}

Differentiating Eq.~(\ref{eq:w-deficit-transform}) and using Eq.~(\ref{eq:w-state-ode}) gives the attached-flow deficit-state equation
\begin{equation}
  \dot{r}_i^w
  +
  \frac{\beta_i}{T_0}r_i^w
  =
  A_i
  \dot{\widetilde{w}}_{3/4}.
  \label{eq:w-deficit-ode}
\end{equation}
The term $\beta_i r_i^w/T_0$ describes the decay of the stored state, whereas $A_i\dot{\widetilde{w}}_{3/4}$ represents the new contribution to the wake-memory. Together with Eqs.~(\ref{eq:wE-deficit-states}) and (\ref{eq:alphaE-zero-lift-shift}), Eq.~(\ref{eq:w-deficit-ode}) is equivalent to the direct downwash-state formulation.

When the deficit-state formulation is embedded in a full dynamic-stall model, a scaling of the newly generated wake memory term with the separation state has been shown to provide a reasonable engineering treatment for issues arising during rapid angle-of-attack variations, such as those encountered in vertical-axis wind turbine applications \citep{Pirrung2018}.
The corresponding extension and its time-discrete implementation are described in Appendix~\ref{app:indicial-lag-updates}.

The preceding subsections provide the classical sectional loads in the chordwise normal and tangential directions together with the attached-flow shed wake memory.
Because the normal and tangential directions are fixed relative to the airfoil chord, this chord-based force representation is unambiguous.
For coupling with the 2-D airfoil polar and with the broader dynamic stall formulation, it is convenient to express the sectional forces in terms of lift and drag.
The 2-D airfoil polar is conventionally provided in lift and drag coefficients, and the separation related terms in the full dynamic stall model are also formulated in terms of lift and drag.
Unlike the chord-based normal and tangential components, however, lift and drag must be defined relative to a local reference flow direction.
Specifying this reference direction is therefore a key, yet often overlooked, aspect of a consistent generalized lifting-line formulation.
The following subsections develop this formulation by relating the blade and wake model to the 2-D section, interpreting the circulatory force, coupling the unsteady formulation with the airfoil polar, and projecting the resulting force contributions relative to the selected flow direction.

\subsection{Coupling to the blade and wake model}
\label{sec:blade-wake-coupling}

In generalized lifting-line methods, including BEM, LL, and AL, the outer 3-D rotor and wake problem combines the onset flow, blade motion, and induced velocity to determine the local relative velocity vector at each aerodynamic calculation point.
This velocity vector is projected onto the 2-D airfoil-section plane and supplied as the external flow to the inner 2-D airfoil problem \citep{Li2022_How}.
Following the input--output machine interpretation introduced in Sect.~\ref{sec:dimensional-thin-airfoil-loads}, the outer flow solution and sectional kinematics together provide the relative flow and kinematic histories required by the inner 2-D model, including the information needed to determine the chordwise downwash distribution and its time variation.
The inner model combines the 2-D airfoil polar with the non-circulatory force and moment contributions derived in the present work and the attached-flow indicial wake memory. The indicial wake memory is included only where the shed wake dynamics are not represented by the outer rotor and wake model.
The projected flow defines the relative streamwise velocity $U$ and the angle of attack $\alpha_{\chi}$ at the calculation point.

Let $\chi$ denote the location of the aerodynamic calculation point as a fraction of the chord measured from the leading edge, with $\chi=0$ at the leading edge and $\chi=1$ at the trailing edge.
The velocity relative to the airfoil due to its translational motion is uniform over the chord.
The treatment of the trailed wake induced velocity supplied by the outer 3-D wake model, hereafter referred to as induction, depends on the method used.
In BEM and conventional LL and AL implementations, this induction is evaluated at a single aerodynamic calculation point and is therefore treated as uniform over the chord when coupled to the inner 2-D airfoil model.
Multi-point LL formulations may instead provide the induction at more than one chordwise location, for example at the quarter- and three-quarter-chord points in \citet{Gaunaa2026}, thereby retaining its chordwise variation.

Following \citet{Hansen2004}, variations in the incoming flow, including gusts, are treated as instantaneous changes over the entire airfoil section.
For the instantaneous sectional relative flow velocity, changes due to translational airfoil motion and variations in the incoming flow therefore need not be distinguished at the sectional aerodynamic level.
The accuracy of this approximation for unsteady inflow was investigated by \citet{buhl2005a}, and its application in aeroelastic wind turbine simulations is discussed further by \citet{Li2022_How}.
In contrast, the effective torsional rate $\dot{\theta}$ is treated separately because it induces a linear variation along the chord in the velocity component normal to the airfoil and, consequently, in the local angle of attack.
This kinematic variation in the 2-D airfoil model is independent of the treatment of the outer trailed wake induced velocity.

For a wind turbine blade section, the effective torsional rate may result from prescribed blade pitching, blade torsional deformation, or the projection of the rotor angular velocity onto the 2-D airfoil-section plane, as occurs for coned HAWTs and VAWTs \citep{Pirrung2018,Li2022_How,Li2025_Disentangle}.
The aerodynamic formulation depends only on the resulting effective torsional rate and does not distinguish between its origins.

For a rigid airfoil section, the effective torsional rate causes the normal velocity, and hence the local angle of attack, to vary linearly along the chord.
The angles of attack at the quarter-chord and three-quarter-chord points are therefore reconstructed as
\begin{gather}
  \alpha_{1/4}
  =
  \alpha_{\chi}
  +
  \left(\frac{1}{4}-\chi\right)
  \frac{c\dot{\theta}}{U},
  \label{eq:calc-point-quarter-chord-angle}\\
  \alpha_{3/4}
  =
  \alpha_{\chi}
  +
  \left(\frac{3}{4}-\chi\right)
  \frac{c\dot{\theta}}{U}.
  \label{eq:calc-point-three-quarter-chord-angle}
\end{gather}
Their difference is independent of the location of the aerodynamic calculation point:
\begin{equation}
  \alpha_{3/4}-\alpha_{1/4}
  =
  \frac{b\dot{\theta}}{U}
  =
  T_0\dot{\theta}.
  \label{eq:quarter-three-quarter-difference}
\end{equation}

For example, in the aerodynamic module of HAWC2, the aerodynamic calculation point is located at the three-quarter-chord point~\citep{hawc2manual}.
Therefore,
\begin{gather}
  \alpha_{\chi}=\alpha_{3/4},
  \label{eq:hawc2-calculation-point-angle}\\
  \alpha_{1/4}
  =
  \alpha_{3/4}-T_0\dot{\theta}.
  \label{eq:hawc2-quarter-chord-angle}
\end{gather}

The mid-chord heaving acceleration $\ddot{\epsilon}$ represents the kinematic contribution from airfoil section motion to the time variation of the relative flow normal to the airfoil chord at mid-chord.
In an aeroelastic model, the kinematic contribution can arise from structural deformation and rigid body rotor motion.
The chord-normal relative velocity may also vary as the blade moves through a spatially and temporally varying flow field, including gusts and turbulence.
\footnote{In the present implementation of the HAWC2 aerodynamic module, the contribution from temporal variation of the chord-normal relative velocity to the non-circulatory mid-chord acceleration term is neglected.}
The kinematic contributions from the blade and rotor motion must nevertheless be included consistently.
It is important to notice that the contributions from the projection of the centripetal acceleration due to rotation of a deflected or coned rotor into the airfoil section must be included in this term. An extreme case in this respect occurs for vertical-axis wind turbines (VAWTs), where the projection of the centripetal acceleration onto the airfoil normal direction can be significant.

The treatment of the shed wake memory depends on the outer aerodynamic model.
In BEM models, the outer induction model based on momentum theory does not explicitly model the shed wake from the blades. The attached flow indicial states are therefore used to represent the local 2-D shed wake memory and determine $\widetilde{w}_E$ and $\alpha_E$.
Conversely, when a lifting-line (LL) or actuator-line (AL) solver explicitly resolves the shed wake induction in the outer wake or flow solution, this contribution is already included in the local relative velocity and therefore in $\alpha_{3/4}$ supplied to the 2-D airfoil model.
To avoid double counting, the attached flow indicial states are then omitted by setting $\alpha_E=\alpha_{3/4}$.


\subsection{Circulatory force magnitude and direction}
\label{sec:circulatory-force}

First, the general relation between the chord-based normal and tangential forces and the lift and drag components is introduced.

\begin{figure}[!htbp]
  \centering
  \includegraphics[width=0.5\linewidth]{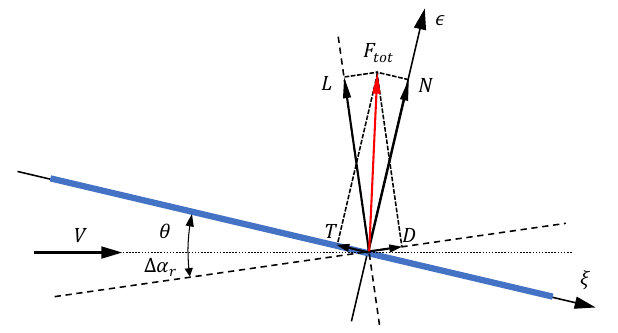}
  \caption{Projection of the normal and tangential force components into lift and drag defined relative to a selected flow direction. The selected direction differs from the onset-flow direction by $\Delta\alpha_r$, such that $\alpha_r=\theta+\Delta\alpha_r$. Adapted from \citet{Li2022_How} with permission.}
  \label{fig:force-projection}
\end{figure}

Let $\Delta\alpha_r$ denote the signed angle between the onset-flow direction and a selected reference flow direction.
The angle between the chord and the selected reference flow direction is then $\alpha_r=\theta+\Delta\alpha_r$, as illustrated in Fig.~\ref{fig:force-projection}.

The chord-axis force coefficients are transformed to lift and drag relative to this direction according to
\begin{gather}
  C_{l,r}
  =
  C_N\cos\alpha_r
  +
  C_T\sin\alpha_r,
  \label{eq:rotate-lift}
  \\
  C_{d,r}
  =
  C_N\sin\alpha_r
  -
  C_T\cos\alpha_r.
  \label{eq:rotate-drag}
\end{gather}

Under the small-angle approximation used in classical thin-airfoil theory,
\begin{gather}
  C_{l,r}
  \approx
  C_N,
  \label{eq:rotate-lift-small-angle}
  \\
  C_{d,r}
  \approx
  C_N\alpha_r
  -
  C_T.
  \label{eq:rotate-drag-small-angle}
\end{gather}

The circulatory-force relations developed below retain the small-angle approximation of classical thin-airfoil theory.
The non-circulatory normal and tangential force contributions are later projected using the full trigonometric relations in Eqs.~(\ref{eq:rotate-lift}) and (\ref{eq:rotate-drag}).
The pitching moment is defined about the quarter-chord point and is unchanged by the force rotation.

Analysis of the unsteady thin airfoil equations identifies distinct roles for the three-quarter-chord and quarter-chord flow quantities \citep{Oye1981,Gaunaa2010,Bergami2012ATEFlap}.
The three-quarter-chord quantity governs the circulatory lift magnitude, while the quarter-chord flow direction provides the reference direction for resolving the circulatory lift and drag \citep{Li2022_How}.

The zero-lift-shifted three-quarter-chord downwash $\widetilde{w}_{3/4}$ is the input to the 2-D shed wake model in the present formulation.
The indicial-response states account for the history of this input and determine the zero-lift-shifted equivalent wake-lagged downwash $\widetilde{w}_E$, from which the effective angle $\alpha_E$ is recovered.
The three-quarter-chord angle $\alpha_{3/4}$ defines the quasi-steady target of the circulatory response, while $\alpha_E$ represents the response after accounting for the shed wake memory.
In classical linear thin airfoil theory for a symmetric airfoil, the corresponding circulatory normal-force coefficient is
\begin{equation}
  C_{N,\mathrm{circ}}
  =
  2\pi\alpha_E.
  \label{eq:cn-circ-linear}
\end{equation}
Thus, $\alpha_E$ determines the circulatory lift magnitude.

The geometric quarter-chord flow direction $\alpha_{1/4}$ has a different role and is used as the reference direction for resolving the circulatory lift and drag.
Using the kinematic relation $\alpha_{1/4}=\alpha_{3/4}-T_0\dot{\theta}$, the circulatory normal-force coefficient and the circulatory tangential-force contribution in Eq.~(\ref{eq:ct}) give
\begin{equation}
  C_{d,\mathrm{ind}}
  =
  2\pi\alpha_E\alpha_{1/4}
  -
  C_{T,\mathrm{circ}}^{\mathrm{suction}}
  =
  2\pi\alpha_E
  \left(
    \alpha_{3/4}-\alpha_E
  \right).
  \label{eq:quarter-chord-drag-balance}
\end{equation}
Here, $C_{d,\mathrm{ind}}$ is the circulatory drag coefficient resolved relative to the quarter-chord flow direction.

In the quasi-steady limit, $\alpha_E=\alpha_{3/4}$, and the induced drag in Eq.~(\ref{eq:quarter-chord-drag-balance}) vanishes.
The integrated circulatory normal force and leading-edge suction therefore produce zero drag when resolved relative to the quarter-chord flow direction.
This property motivates the use of $\alpha_{1/4}$ as the reference direction for the circulatory lift and drag.

Under unsteady conditions, the shed wake memory causes $\alpha_E$ to differ from $\alpha_{3/4}$.
The integrated circulatory force then has a non-zero drag component when resolved relative to $\alpha_{1/4}$, as given by Eq.~(\ref{eq:quarter-chord-drag-balance}).

The direction of the circulatory-force resultant can be identified as the direction in which its drag component vanishes.
The angle of the circulatory-force resultant relative to the chord is denoted by $\alpha_{\mathrm{circ,dir}}$.
Using the circulatory tangential-force contribution in Eq.~(\ref{eq:ct}), this angle is
\begin{equation}
  \alpha_{\mathrm{circ,dir}}
  =
  \alpha_E-T_0\dot{\theta}
  =
  \alpha_{1/4}
  -
  \left(
    \alpha_{3/4}-\alpha_E
  \right).
  \label{eq:circulatory-force-direction}
\end{equation}
The shed wake lag $\alpha_{3/4}-\alpha_E$ therefore shifts the circulatory-force direction away from the geometric quarter-chord flow direction.
Resolving this force relative to $\alpha_{1/4}$ produces the induced drag in Eq.~(\ref{eq:quarter-chord-drag-balance}).

\subsection{Coupling with the 2-D airfoil polar}
\label{sec:polar-coupling}

In the classical thin airfoil formulation, the circulatory lift magnitude is governed by the effective angle of attack $\alpha_E$ through a linear flat plate relation with a lift-curve slope of $2\pi$.
To represent the aerodynamic characteristics of a real airfoil, this linear relation is replaced by a 2-D airfoil polar obtained from measurements or CFD simulations \citep{Hansen2004}.

The polar is evaluated at $\alpha_E$ and provides
\begin{gather}
  C_{l,\mathrm{circ}}^{\mathrm{dyn}}
  \equiv
  C_l^{\mathrm{polar}}(\alpha_E),
  \label{eq:polar-circulatory-lift}\\
  C_d^{\mathrm{profile}}
  \equiv
  C_d^{\mathrm{polar}}(\alpha_E),
  \label{eq:polar-profile-drag}\\
  C_m^{\mathrm{profile}}
  \equiv
  C_{m,1/4}^{\mathrm{polar}}(\alpha_E).
  \label{eq:polar-profile-moment}
\end{gather}
Here, the superscript $\mathrm{dyn}$ denotes the unsteady circulatory lift response, and $C_{l,\mathrm{circ}}^{\mathrm{dyn}}$ is obtained by evaluating the polar at the wake-lagged effective angle $\alpha_E$.

In the present formulation, the polar lookup provides the circulatory lift magnitude, profile drag, and quarter-chord pitching moment.
The airfoil polar accounts for the effects of airfoil camber and viscosity on these quantities.
The non-circulatory loads and the induced drag associated with the unsteady 2-D shed wake are added separately.

The corresponding 2-D shed wake-induced drag is approximated following the formulations of \citet{Hansen2004} and \citet{Bergami2012ATEFlap} as
\begin{equation}
  C_{d,\mathrm{ind}}
  =
  C_{l,\mathrm{circ}}^{\mathrm{dyn}}
  \left(
    \alpha_{3/4}-\alpha_E
  \right).
  \label{eq:cdind-alpha}
\end{equation}

\subsection{Force projection onto the selected flow direction}
\label{sec:force-projection}

When lift and drag are defined relative to the local flow direction at the aerodynamic calculation point $\chi$, the reference direction is $\alpha_r=\alpha_{\chi}$. The circulatory force is therefore decomposed into lift and drag components, which are normal and tangential to $\alpha_{\chi}$, respectively. For convenience, the tangential projection is expressed using the angular difference between $\alpha_{\chi}$ and the zero-circulatory-drag direction $\alpha_{\mathrm{circ,dir}}$:
\begin{equation}
  C_{d,\mathrm{circ},\chi}
  =
  C_{l,\mathrm{circ}}^{\mathrm{dyn}}
  \sin
  \left[
    \left(
      \alpha_{3/4}
      -
      \alpha_E
    \right)
    +
    \left(
      \chi-\frac{1}{4}
    \right)
    \frac{c\dot{\theta}}{U}
  \right].
  \label{eq:calc-point-circulatory-drag-exact}
\end{equation}
Using the small-angle approximation and the definition of $C_{d,\mathrm{ind}}$, this becomes
\begin{equation}
  C_{d,\mathrm{circ},\chi}
  \approx
  C_{d,\mathrm{ind}}
  +
  C_{l,\mathrm{circ}}^{\mathrm{dyn}}
  \left(
    \chi-\frac{1}{4}
  \right)
  \frac{c\dot{\theta}}{U}.
  \label{eq:calc-point-drag}
\end{equation}

When the reference direction is the geometric quarter-chord flow direction, $\chi=1/4$, the second term in Eq.~(\ref{eq:calc-point-drag}) vanishes, and the circulatory drag reduces to $C_{d,\mathrm{ind}}$.

When the reference direction is the three-quarter-chord flow direction, as in HAWC2 \citep{Madsen2020BEM}, $\chi=3/4$.
Lift and drag are then defined relative to the local flow direction at the three-quarter-chord point.
Equation~(\ref{eq:calc-point-drag}) becomes
\begin{equation}
  C_{d,\mathrm{circ},3/4}
  \approx
  C_{d,\mathrm{ind}}
  +
  C_{l,\mathrm{circ}}^{\mathrm{dyn}}
  T_0\dot{\theta}.
  \label{eq:cddir}
\end{equation}
Although the two reference directions lead to different lift and drag decompositions, they describe the same aerodynamic force when the corresponding projection is implemented consistently.

For the three-quarter-chord reference direction, the second term in Eq.~(\ref{eq:cddir}) is denoted by $C_d^{\Delta\alpha}=C_{l,\mathrm{circ}}^{\mathrm{dyn}}T_0\dot{\theta}$ and represents the direction-projection contribution. It is retained separately from the 2-D shed wake-induced drag $C_{d,\mathrm{ind}}$. Both circulatory contributions are distinct from the non-circulatory torsional-rate-squared contribution $C_{T,\mathrm{NC}}^{\dot{\theta}^2}$ in Eq.~(\ref{eq:ct}), which is quadratic in the torsional rate and does not depend on the circulatory lift coefficient.

\subsection{Final implementation form}
\label{sec:implementation-form}

The general formulation can be implemented using different aerodynamic calculation points and force reference directions.
The convention used in HAWC2 version 13.2 is presented here as one possible approach.
The three-quarter-chord point is used as the aerodynamic calculation point, and lift and drag are defined relative to the local flow direction at that point.

For each blade section, a local sectional coordinate system $S$ is defined following the HAWC2 convention. Its origin is located at the airfoil mid-chord point, the sectional $x$-axis lies along the chord and is positive from the trailing edge towards the leading edge, and the sectional $y$-axis is normal to the chord and positive from the pressure side towards the suction side. The sectional $z$-axis completes the right-handed coordinate system. The orientation of the sectional coordinate system follows the instantaneous blade geometry, such that prebend, blade deflection, twist, and torsional deformation are included. Translational and rotational kinematic quantities provided by the structural solver in the global coordinate system $G$ are transformed to the sectional coordinate system using $\mathbf{T}_{G\rightarrow S}$.

For each blade section, the outer solver provides the relative streamwise velocity $U$ and the angle of attack $\alpha_{\chi}$ at the aerodynamic calculation point.
Incoming-flow variations, including turbulence, therefore enter $U$ and $\alpha_{\chi}$ through the instantaneous relative flow.
The sectional kinematics provide the effective torsional rate $\dot{\theta}$ and the mid-chord heaving acceleration $\ddot{\epsilon}$, as described in Sect.~\ref{sec:blade-wake-coupling}.
In the present HAWC2 implementation, temporal variations of the flow field are included in the instantaneous relative velocity and angle of attack, but their contributions to the rate of change of the relative flow magnitude and direction are neglected.
A fully consistent treatment of these contributions in aeroelastic simulations with turbulent inflow is beyond the scope of the present work and is left for future work.
For the HAWC2 implementation, the structural kinematics are evaluated at the three-quarter-chord point. Denoting its chord-normal acceleration in sectional coordinates by $a_{3/4,y}^{S}$, the mid-chord heaving acceleration can be expressed as
\begin{equation}
    \ddot{\epsilon}
    =
    a_{3/4,y}^{S}
    +
    \frac{c}{4}\dot{\omega}_{z}^{S}
    +
    \frac{c}{4}\omega_x^{S}\omega_y^{S}
    \approx
    a_{3/4,y}^{S}
    +
    \frac{c}{4}\dot{\omega}_{z}^{S}.
  \label{eq:epsilon-from-three-quarter-hawc2}
\end{equation}
In the present HAWC2 implementation, the approximated form in Eq.~(\ref{eq:epsilon-from-three-quarter-hawc2}) is used. The three-dimensional quadratic rotational contribution $\frac{c}{4}\omega_x^{S}\omega_y^{S}$, associated with the point transformation, is neglected.

The effective torsional rate $\dot{\theta}$ is correspondingly the angular-velocity component about the sectional $z$-axis.
For the present choice $\chi=3/4$, $\alpha_{\chi}=\alpha_{3/4}$.
The corresponding zero-lift-shifted downwash $\widetilde{w}_{3/4}=U(\alpha_{3/4}-\alpha_0)$ is used to update the time-domain 2-D shed wake states described in Sect.~\ref{sec:wake-state}, which determine $\widetilde{w}_E$ and the effective angle of attack $\alpha_E=\alpha_0+\widetilde{w}_E/U$.

With these quantities, the total lift coefficient is
\begin{equation}
  C_{l,3/4}^{\mathrm{tot}}
  =
  C_{l,\mathrm{circ}}^{\mathrm{dyn}}
  +
  C_{N,\mathrm{NC}}
  \cos\alpha_{3/4}
  +
  C_{T,\mathrm{NC}}^{\dot{\theta}^2}
  \sin\alpha_{3/4}.
  \label{eq:cltot}
\end{equation}
Here, $C_{l,\mathrm{circ}}^{\mathrm{dyn}}$ is obtained by evaluating the airfoil polar at $\alpha_E$ using Eq.~(\ref{eq:polar-circulatory-lift}).
The non-circulatory coefficients $C_{N,\mathrm{NC}}$ and $C_{T,\mathrm{NC}}^{\dot{\theta}^2}$ are given by Eqs.~(\ref{eq:cn}) and (\ref{eq:ct}), respectively, and their lift components follow from the projection in Eq.~(\ref{eq:rotate-lift}).

The total drag coefficient is
\begin{equation}
  C_{d,3/4}^{\mathrm{tot}}
  =
  C_d^{\mathrm{profile}}
  +
  C_{d,\mathrm{ind}}
  +
  C_d^{\Delta\alpha}
  +
  C_{N,\mathrm{NC}}
  \sin\alpha_{3/4}
  -
  C_{T,\mathrm{NC}}^{\dot{\theta}^2}
  \cos\alpha_{3/4}.
  \label{eq:cdtot}
\end{equation}
Here, $C_d^{\mathrm{profile}}$ is obtained from the airfoil polar at $\alpha_E$ using Eq.~(\ref{eq:polar-profile-drag}), and the 2-D shed wake-induced drag $C_{d,\mathrm{ind}}$ is defined in Eq.~(\ref{eq:cdind-alpha}).
The direction-projection contribution $C_d^{\Delta\alpha}$ is defined by the second term in Eq.~(\ref{eq:cddir}).
The non-circulatory drag components follow from the projection in Eq.~(\ref{eq:rotate-drag}).

The full non-circulatory quarter-chord pitching moment is given by Eq.~(\ref{eq:cm-full}) and is repeated here for completeness:
\begin{equation}
  \begin{split}
    C_{m,\mathrm{NC}}^{\mathrm{full}}
    &=
    \frac{1}{4}\pi T_0
    \frac{\ddot{\epsilon}}{U}
    -
    \frac{1}{2}\pi T_0\dot{\theta}
    -
    \frac{1}{16}\pi T_0^2\ddot{\theta}
    \\
    &=
    -\frac{1}{4}
    C_{N,\mathrm{NC}}^{\mathrm{acc}}
    -
    \frac{1}{2}
    C_{N,\mathrm{NC}}^{\mathrm{tor}}
    -
    \frac{1}{16}\pi T_0^2\ddot{\theta}.
  \end{split}
  \label{eq:cm-full-repeated}
\end{equation}
In HAWC2 version 13.2, only the final term, which is proportional to the torsional acceleration $\ddot{\theta}$, is neglected. Accordingly, $C_{m,\mathrm{NC}}$ denotes the remaining non-circulatory pitching moment in the following implementation form:
\begin{equation}
  C_{m,1/4}^{\mathrm{tot}}
  =
  C_m^{\mathrm{profile}}
  +
  C_{m,\mathrm{NC}},
  \label{eq:cmtot}
\end{equation}
where $C_m^{\mathrm{profile}}$ is obtained from the airfoil polar at $\alpha_E$ using Eq.~(\ref{eq:polar-profile-moment}).
Because the non-circulatory added mass terms depend directly on the acceleration of the blade section, explicit aeroelastic coupling can become numerically unstable.
In HAWC2 version 13.2, the mid-chord heaving acceleration from the converged blade kinematics at the previous time step is used, which is therefore kept fixed during the aeroelastic iterations.
An alternative is to include the aerodynamic added mass in the blade structural mass matrix.

If lift and drag are defined relative to a local flow direction other than that at the three-quarter-chord point, the force coefficients can be transformed using Eqs.~(\ref{eq:rotate-lift}) and (\ref{eq:rotate-drag}).


The force application point is a separate choice from the aerodynamic calculation point and the reference direction for lift and drag. In the HAWC2 implementation, the sectional force is applied at the three-quarter-chord point, whereas the aerodynamic pitching moment in Eq.~(\ref{eq:cmtot}) is defined about the quarter-chord point. After conversion to dimensional sectional loads, static equivalence is preserved by transferring the moment according to
\begin{equation}
  \mathbf{M}_{3/4}^{\mathrm{tot}}
  =
  \mathbf{M}_{1/4}^{\mathrm{tot}}
  +
  \left(
    \mathbf{r}_{1/4}
    -
    \mathbf{r}_{3/4}
  \right)
  \times
  \mathbf{F},
  \label{eq:quarter-three-quarter-moment-shift}
\end{equation}
where $\mathbf{F}$ is the dimensional sectional aerodynamic force. The moment-transfer term acts about the local spanwise axis and therefore contributes directly to the blade torsional moment. For a straight blade lying in the rotor plane, it contributes only to the blade torsional moment. Coning, prebend, or other out-of-plane geometry tilts the local spanwise axis, causing part of the transferred moment to project onto the rotor axis and contribute to the aerodynamic rotor torque.

Algorithm~\ref{alg:attached-flow-implementation} summarizes the resulting attached-flow update within an iterative aeroelastic simulation using a BEM aerodynamic model, following the conventions and implementation sequence used in the HAWC2 code. Quantities required from the previous time step remain fixed during the iterations of the new time step, while the current flow and sectional kinematics are updated in each iteration.

\begin{algorithm}[!htbp]
\caption{Evaluation of the sectional attached-flow aerodynamic model within an iterative aeroelastic simulation using a BEM aerodynamic model}
\label{alg:attached-flow-implementation}
\begin{algorithmic}
\FOR{each time step $t_{n+1}$}
\FOR{each aeroelastic iteration}
\FOR{each blade section}
\IF{this is the first iteration at $t_{n+1}$}
\STATE Save the converged aerodynamic state variables from the previous time step $t_n$ for use in all iterations at this time step.
\ENDIF
\STATE \textit{Construct the input to the sectional 2-D model.}
\STATE Obtain the current blade section kinematics from the structural solver.
\STATE Obtain the local inflow velocity, including shear and turbulence.
\STATE Determine the relative flow velocity from the inflow, induced velocity, and blade motion.
\STATE Determine $U$ and $\alpha_{3/4}$ from the relative flow velocity in the sectional coordinate system $S$, using components in the airfoil plane.
\STATE Determine $\dot{\theta}$ from the sectional kinematics and compute $\ddot{\epsilon}$ from Eq.~(\ref{eq:epsilon-from-three-quarter-hawc2}).
\STATE \textit{Evaluate the unsteady sectional 2-D model.}
\STATE Evaluate the non-circulatory terms using $\ddot{\epsilon}_n$ from the previous time step and $\dot{\theta}$ from the current time step.
\STATE Compute $\widetilde{w}_{3/4}=U(\alpha_{3/4}-\alpha_0)$ and update the shed-wake states using Eq.~(\ref{eq:appendix-deficit-physical-time-step}) with $f_{\mathrm{scale}}=1$.
\STATE Compute $\widetilde{w}_E$ and $\alpha_E$ using Eq.~(\ref{eq:appendix-deficit-alphaE}) and evaluate the airfoil polar at $\alpha_E$ using Eqs.~(\ref{eq:polar-circulatory-lift}) to (\ref{eq:polar-profile-moment}).
\STATE Compute $C_{d,\mathrm{ind}}$ and $C_d^{\Delta\alpha}$ using Eqs.~(\ref{eq:cdind-alpha}) and (\ref{eq:cddir}).
\STATE Assemble $C_{l,3/4}^{\mathrm{tot}}$, $C_{d,3/4}^{\mathrm{tot}}$, and $C_{m,1/4}^{\mathrm{tot}}$ using Eqs.~(\ref{eq:cltot}), (\ref{eq:cdtot}), and (\ref{eq:cmtot}).
\STATE \textit{Return the sectional aerodynamic output to the rotor model.}
\STATE Convert the coefficients to dimensional sectional loads, transfer the loads to the selected reference point, and supply them to the structural solver.
\ENDFOR
\ENDFOR
\ENDFOR
\end{algorithmic}
\end{algorithm}

\subsubsection{Practical interpretation of sectional coefficients}
\label{sec:coefficient-interpretation}

The sectional coefficients require careful interpretation when the relative velocity is small or when lift and drag are defined relative to the three-quarter-chord flow direction.
At small relative velocities, the dynamic pressure becomes small and the non-circulatory lift coefficients can therefore become very large, while the associated dimensional forces remain bounded and physically meaningful~\citep{hawc2manual}.
The drag coefficient may also become negative because the selected reference direction changes the decomposition of the sectional force into lift and drag.
Such coefficient values do not necessarily indicate an incorrect aerodynamic force, and the corresponding dimensional force components should therefore also be examined.

\section{Rotor-level implementation benchmark and verification cases}
\label{sec:verification}

The classical sectional pitching and heaving responses have been documented and verified in previous formulations of the attached-flow model \citep{Hansen2004,Gaunaa2010,Bergami2012ATEFlap}.
The purpose of the following cases is therefore not to revalidate the underlying 2-D theory, but to verify its implementation in wind turbine aerodynamic solvers.
For wind turbine blades, the rotor 3-D kinematics must be projected into the sectional coordinate system and used to determine the required sectional flow quantities, while the resulting forces and moments must be transformed back to the rotor reference system.

The two rotor-level cases are used for different verification purposes.
The coned straight blade case provides a cross-method numerical benchmark and assesses how omitting selected terms changes the predicted rotor-integrated thrust and power.
The zero-onset VAWT case provides an analytical consistency test by extending the studies of \citet{Pirrung2018} and \citet{Li2022_How} to include all circulatory and non-circulatory contributions in the ideal thin-airfoil limit.
The resulting torque contributions are shown to cancel, giving zero total aerodynamic torque.

\subsection{Coned straight blade case}
\label{sec:case-coned}

The coned straight blade case provides a controlled rotor-level test of the coupling between the 3-D rotor kinematics and the unsteady 2-D airfoil model.
At non-zero cone angles, effective sectional torsional rate and acceleration contributions arise even under steady operating conditions, allowing their influence on the predicted rotor loads to be isolated and quantified.

The effective sectional torsional rate accounts for the variation in local flow angle along the chord and is used to determine the flow angles at the three-quarter- and quarter-chord points from the flow evaluated at a single aerodynamic calculation point.\footnote{This procedure is referred to as the \emph{one-point correction} by \citet{Li2022_How}.}
\citet{Li2022_How} examined its influence primarily through comparisons of sectional blade loads.
These comparisons are not repeated here; instead, the present analysis considers rotor-integrated quantities over a wide range of cone angles.

The blade-element momentum (BEM), lifting-line (LL), and computational fluid dynamics (CFD) calculations provide a rotor-level numerical benchmark across aerodynamic solvers of increasing fidelity.
In addition, three BEM calculations are performed, each omitting one contribution from the complete formulation: the direction-projection contribution $C_d^{\Delta\alpha}$, the non-circulatory torsional-rate-squared tangential-force contribution $C_{T,\mathrm{NC}}^{\dot{\theta}^2}$, or the non-circulatory force and moment contributions associated with mid-chord heaving acceleration.

\subsubsection{Test configuration and objective}

The present study builds on the comparison used in \citet{Li2022_How}, but uses an updated straight-bladed configuration based on the IEA 10 MW reference wind turbine \citep{IEA37} and a different operating condition.
All CFD, LL, and BEM results presented here are obtained from new calculations.
The blade geometry is prescribed in all simulations, elastic blade deformations are not included, and the blade pitch angle is $0^\circ$.

The rotor tilt is first removed.
The prebend and sweep of the reference blade are then removed so that the half-chord line is straight.
The baseline rotor radius is $R_0=99.555\,\mathrm{m}$, including the hub radius $r_{\mathrm{hub}}=2.8\,\mathrm{m}$.
Coned rotors are generated by rotating each blade about its root connection to the hub, with the cone angle $\kappa$ ranging from $-30^\circ$ to $30^\circ$.
Positive cone angles denote upwind coning.

The rotor thrust and power coefficients are normalized using the projected rotor area
\begin{equation}
  A_{\mathrm{proj}}(\kappa)
  =
  \pi
  \left[
    r_{\mathrm{hub}}
    +
    \left(
      R_0-r_{\mathrm{hub}}
    \right)
    \cos\kappa
  \right]^2.
  \label{eq:coned-projected-area}
\end{equation}
The rotor thrust and power coefficients are calculated from the rotor-integrated thrust and power, $T_{\mathrm{rot}}$ and $P_{\mathrm{rot}}$, respectively, and are defined as
\begin{gather}
  C_{\mathrm{T,rot}}
  =
  \frac{T_{\mathrm{rot}}}
  {\tfrac{1}{2}\rho U_0^2 A_{\mathrm{proj}}(\kappa)},
  \label{eq:coned-thrust-coefficient}\\
  C_{\mathrm{P,rot}}
  =
  \frac{P_{\mathrm{rot}}}
  {\tfrac{1}{2}\rho U_0^3 A_{\mathrm{proj}}(\kappa)}.
  \label{eq:coned-power-coefficient}
\end{gather}

The rotor operates in uniform inflow with no wind shear or yaw misalignment.
The wind speed is $U_0=8\,\mathrm{m\,s^{-1}}$, and the rotational speed is fixed at $\Omega=0.72322\,\mathrm{rad\,s^{-1}}$.
For the baseline rotor with radius $R_0$, the tip-speed ratio is $\lambda_0=\Omega R_0/U_0=9$.
Fully turbulent airfoil polars obtained from 2-D CFD calculations are used consistently in the LL and BEM simulations.

The high-fidelity reference calculations are performed using EllipSys3D, a pressure-based incompressible three-dimensional computational fluid dynamics (CFD) solver \citep{Michelsen1992,Michelsen1994,Sorensen1995}.
The solver uses a finite-volume discretization of the Reynolds-averaged Navier--Stokes (RANS) equations.
The flow is assumed fully turbulent and modeled using the $k$--$\omega$ shear stress transport (SST) turbulence model \citep{menterkomega}.
Body-fitted, wall-resolved meshes are used following the numerical setup described by \citet{Li2022_How}.

The LL calculations are performed using the free-wake vortex solver MIRAS \citep{Ramos-Garcia_we2016} coupled to HAWC2 for the prescribed blade geometry and rigid-body sectional kinematics.
Structural flexibility is disabled, and no elastic blade deformation is included.
MIRAS evaluates the 3-D induced velocity at the quarter-chord point as part of the outer 3-D free-wake problem.
HAWC2 evaluates the rotational and structural kinematic velocities at the three-quarter-chord point and combines them with the MIRAS induced velocity.
The relative flow direction at the three-quarter-chord point is used to define lift and drag.

The BEM calculations are performed using the HAWC2 BEM module.
Within this module, the sectional relative velocity and the reference direction for lift and drag are evaluated at the three-quarter-chord point.
The momentum-theory induction is determined at the blade-element level and is assumed constant along the chord.
For a straight blade with a cone angle $\kappa$, projecting the rotor angular velocity into the airfoil section results in an effective sectional torsional rate
\begin{equation}
  \dot{\theta}
  =
  -\Omega\sin\kappa.
  \label{eq:coned-effective-torsional-rate}
\end{equation}
According to Eq.~(\ref{eq:quarter-three-quarter-difference}), this torsional rate produces a variation in the flow direction along the chord.
The small chordwise variation in the relative velocity magnitude is neglected, as justified numerically by \citet{Li2022_How}.

Because both LL and BEM define lift and drag relative to the three-quarter-chord flow direction, the direction-projection contribution in Eq.~(\ref{eq:cddir}) has the same formulation in both methods.
The complete LL and BEM formulations retain the non-circulatory force and moment contributions.

For the present steady operating condition, the two-dimensional shed wake-induced drag vanishes, so $C_{d,\mathrm{ind}}=0$.
For the steady coned straight blade configuration, the heaving-acceleration and torsional-rate contributions to $C_{N,\mathrm{NC}}$ in Eq.~(\ref{eq:cn}) cancel \citep{Li2022_How}.
The general drag formulation in Eq.~(\ref{eq:cdtot}) reduces to
\begin{equation}
  C_{d,3/4}^{\mathrm{tot}}
  =
  C_d^{\mathrm{profile}}
  +
  C_d^{\Delta\alpha}
  -
  C_{T,\mathrm{NC}}^{\dot{\theta}^2} \cos\alpha_{3/4}.
  \label{eq:coned-reduced-drag}
\end{equation}

The CFD results are compared with LL and BEM results obtained using the complete formulation.
Three additional BEM calculations quantify the changes caused by omitting individual contributions from the complete formulation.
The first omits the direction-projection contribution $C_d^{\Delta\alpha}$.
The second omits the torsional-rate-squared tangential-force contribution $C_{T,\mathrm{NC}}^{\dot{\theta}^2}$.
The third omits both the non-circulatory normal-force contribution associated with the mid-chord heaving acceleration, $C_{N,\mathrm{NC}}^{\mathrm{acc}}$, and its corresponding quarter-chord moment contribution in Eq.~(\ref{eq:cm-full}).

\subsubsection{Results and discussion}

\begin{figure}[!htbp]
  \centering
  \includegraphics[width=0.75\linewidth]{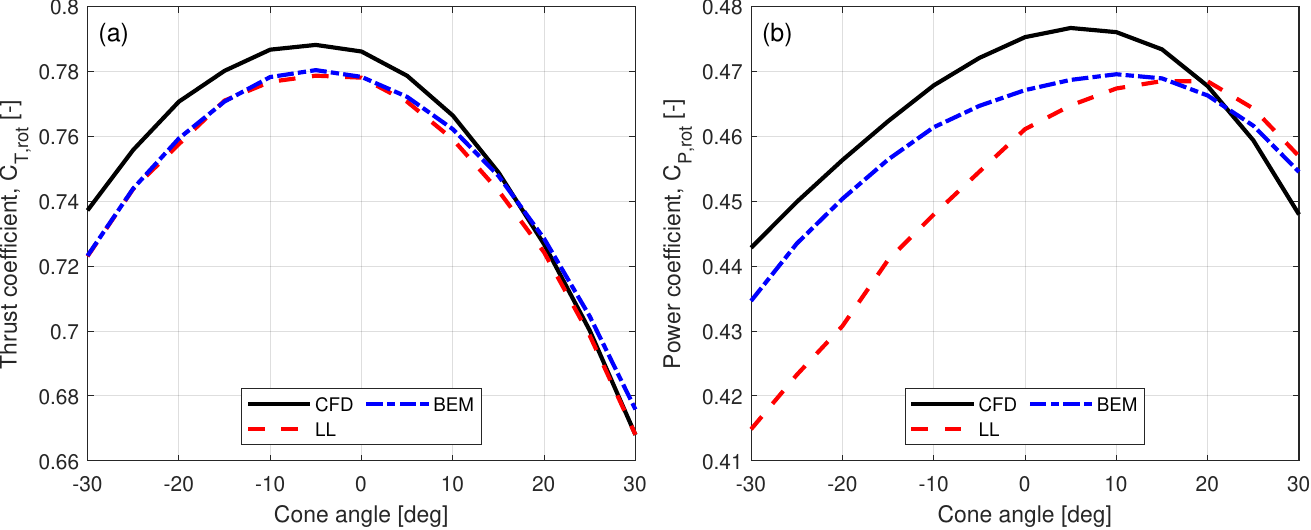}
  \caption{Rotor-integrated (a) thrust coefficient $C_{\mathrm{T,rot}}$ and (b) power coefficient $C_{\mathrm{P,rot}}$ as functions of cone angle for EllipSys3D CFD, MIRAS LL, and the HAWC2 BEM formulation.
  The LL and BEM calculations use fully turbulent airfoil polars.
  Both coefficients are normalized using the cone-angle-dependent projected rotor area $A_{\mathrm{proj}}(\kappa)$.}
  \label{fig:coned-method-comparison}
\end{figure}

Figure~\ref{fig:coned-method-comparison} compares the rotor thrust and power coefficients obtained using CFD, LL, and BEM over the investigated range of cone angles.

The three aerodynamic methods predict similar overall trends, with closer agreement for thrust than for power.
The thrust coefficient reaches a maximum at a slightly negative cone angle and then decreases as the cone angle increases.
The LL and BEM results remain similar, while CFD generally predicts higher thrust coefficients.
The power coefficient also increases and then decreases as the cone angle increases from $-30^\circ$ to $30^\circ$.
CFD predicts its maximum power at a cone angle of approximately $5^\circ$, whereas BEM and LL reach their maximum values at approximately $10^\circ$ and $20^\circ$, respectively.

\begin{figure}[!htbp]
  \centering
  \includegraphics[width=0.75\linewidth]{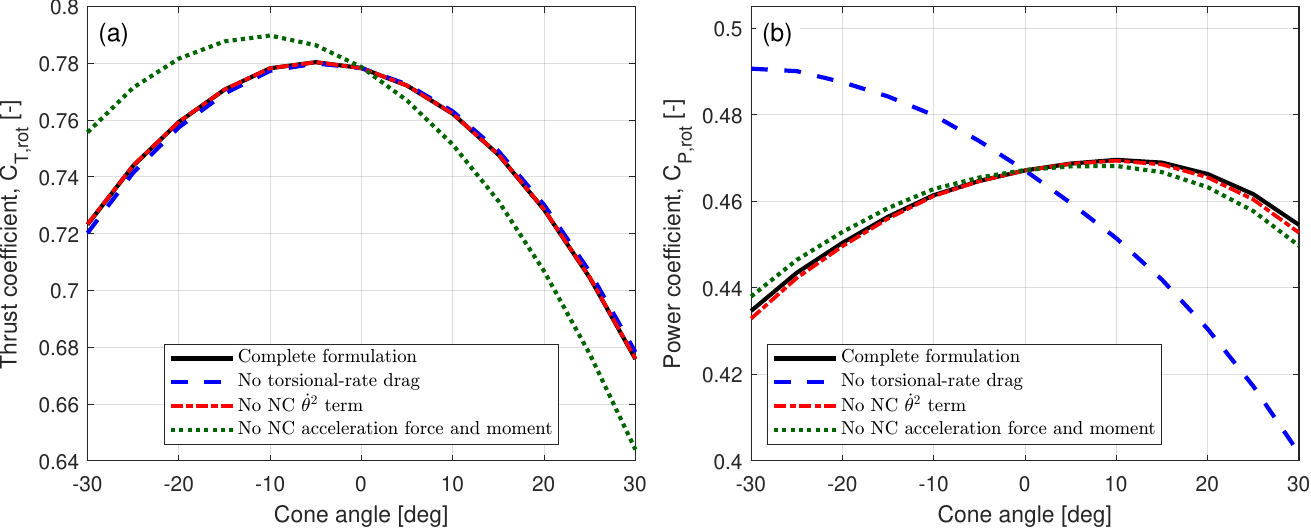}
  \caption{Effect of omitting individual contributions on the rotor-integrated (a) thrust coefficient $C_{\mathrm{T,rot}}$ and (b) power coefficient $C_{\mathrm{P,rot}}$.
  The omitted contributions are the direction-projection term $C_d^{\Delta\alpha}$, the non-circulatory torsional-rate-squared tangential force $C_{T,\mathrm{NC}}^{\dot{\theta}^2}$, and the mid-chord heaving-acceleration force and its associated moment.}
  \label{fig:coned-bem-term-comparison}
\end{figure}

Figure~\ref{fig:coned-bem-term-comparison} compares the complete BEM formulation with the three incomplete variants.

All other aerodynamic inputs and model terms remain unchanged in the three incomplete variants.
Omitting the direction-projection contribution has little influence on thrust but changes the predicted power by approximately $4\,\%$ at $\kappa=\pm10^\circ$, which is significant and consistent with the expected influence of this contribution.
This omission neglects the change in force direction associated with the chordwise variation in angle of attack while retaining the force magnitude determined from the three-quarter-chord aerodynamic state.
The resulting difference primarily affects the tangential loading and rotor-integrated power, whereas its influence on the axial loading and rotor-integrated thrust is smaller, as analysed by \citet{Li2022_How}.

The torsional-rate-squared tangential force has a negligible influence on both thrust and power for the present case.
By contrast, omitting the mid-chord heaving-acceleration contribution primarily affects thrust.
The thrust difference is approximately $1.5\,\%$ at $\kappa=\pm10^\circ$ and increases to about $4.5\,\%$ at $\kappa=\pm30^\circ$, whereas the corresponding power difference remains below approximately $1.1\,\%$.
Omitting the heaving-acceleration contribution removes the cancellation between the two non-circulatory normal-force terms and therefore produces a non-zero non-circulatory normal force.

For the steady coned blade, the heaving-acceleration and torsional-rate contributions to the non-circulatory normal force cancel.
Omitting the heaving-acceleration contribution leaves the torsional-rate contribution unbalanced, producing a spurious non-circulatory normal force and therefore a larger change in thrust.
At zero cone angle, the effective torsional rate vanishes, so the complete formulation and all three incomplete variants give identical results.

\subsection{Zero-onset VAWT: complete torque cancellation in the ideal thin-airfoil limit}
\label{sec:case-vawt}

\subsubsection{Test configuration and objective}

The configuration follows the zero-onset flow vertical-axis wind turbine (VAWT) case considered by \citet{Pirrung2018} and \citet{Li2022_How}.
For each blade, consider an ideal two-dimensional airfoil section that moves on a circular path of radius $R$ with constant angular velocity $\Omega$ and zero onset flow velocity.
The circular motion gives a non-zero sectional relative velocity, effective torsional rate, and centripetal acceleration, producing both circulatory and non-circulatory contributions.

Following \citet{Pirrung2018}, each blade is mounted at the half-chord point for convenience in the derivation, although the analysis can be generalized to other chordwise mounting locations.
The mounting angle is constant, so the section may produce a non-zero quasi-steady circulatory lift.
Profile drag and pitching moment are neglected so that the torque balance contains only the circulatory and non-circulatory contributions from the attached-flow thin-airfoil formulation.
In the ideal thin-airfoil limit, these contributions must cancel exactly in the total aerodynamic torque.

The flow observed by the airfoil section is steady, and the relevant sectional quantities are
\begin{gather}
  U=\Omega R,
  \label{eq:vawt-relative-velocity}\\
  \dot{\theta}=-\Omega,
  \label{eq:vawt-torsional-rate}\\
  \ddot{\epsilon}=-\Omega^2R.
  \label{eq:vawt-heaving-acceleration}
\end{gather}

Previous work showed that the consistently resolved circulatory force produces zero rotor torque in this configuration \citep{Pirrung2018}, and that the heaving-acceleration and torsional-rate contributions to the non-circulatory normal force cancel \citep{Li2022_How}.
The present analysis completes the torque balance by showing that the torque produced by the non-circulatory torsional-rate-squared tangential force is canceled by the non-circulatory pitching moment.
Together, these results account for all circulatory and non-circulatory thin-airfoil contributions.

\subsubsection{Complete torque balance}

The circulatory torque is zero when the circulatory force direction is treated consistently \citep{Pirrung2018}:
\begin{equation}
  Q_{\mathrm{circ}}
  =
  0.
  \label{eq:vawt-circulatory-torque-cancellation}
\end{equation}
The remaining task is to establish the complete non-circulatory torque balance.

For completeness, the non-circulatory normal-force cancellation shown by \citet{Li2022_How} is first rederived using the notation of the present formulation.
Substitution of Eqs.~(\ref{eq:vawt-relative-velocity}) to (\ref{eq:vawt-heaving-acceleration}) into Eqs.~(\ref{eq:cn-acc}) and (\ref{eq:cn-tor}), with $T_0=b/U$, gives
\begin{gather}
  C_{N,\mathrm{NC}}^{\mathrm{acc}}
  =
  \pi\frac{b}{R},
  \label{eq:vawt-normal-force-acceleration}\\
  C_{N,\mathrm{NC}}^{\mathrm{tor}}
  =
  -\pi\frac{b}{R}.
  \label{eq:vawt-normal-force-torsion}
\end{gather}
Hence,
\begin{equation}
  C_{N,\mathrm{NC}}
  =
  C_{N,\mathrm{NC}}^{\mathrm{acc}}
  +
  C_{N,\mathrm{NC}}^{\mathrm{tor}}
  =
  0.
  \label{eq:vawt-normal-force-cancellation}
\end{equation}

Since $\ddot{\theta}=0$, substitution of Eqs.~(\ref{eq:vawt-normal-force-acceleration}) and (\ref{eq:vawt-normal-force-torsion}) into Eq.~(\ref{eq:cm-full}) gives
\begin{equation}
  C_{m,\mathrm{NC}}
  =
  -\frac{1}{4}
  C_{N,\mathrm{NC}}^{\mathrm{acc}}
  -
  \frac{1}{2}
  C_{N,\mathrm{NC}}^{\mathrm{tor}}
  =
  \frac{\pi b}{4R}.
  \label{eq:vawt-non-circulatory-moment}
\end{equation}

Using $T_0=b/U$ together with Eqs.~(\ref{eq:vawt-relative-velocity}) and (\ref{eq:vawt-torsional-rate}), the torsional-rate-squared contribution in Eq.~(\ref{eq:ct}) becomes
\begin{equation}
  C_{T,\mathrm{NC}}^{\dot{\theta}^2}
  =
  \frac{1}{2}\pi
  \left(
    T_0\dot{\theta}
  \right)^2
  =
  \frac{1}{2}\pi
  \left(
    \frac{b}{R}
  \right)^2.
  \label{eq:vawt-tangential-contribution}
\end{equation}

Using $c=2b$ and the dynamic pressure defined in Eq.~(\ref{eq:dynamic-pressure}), the dimensional non-circulatory quarter-chord moment per unit span and tangential force per unit span are
\begin{gather}
  M_{\mathrm{NC}}
  =
  qc^2C_{m,\mathrm{NC}}
  =
  \frac{q\pi b^3}{R},
  \label{eq:vawt-dimensional-moment}\\
  T_{\mathrm{NC}}^{\dot{\theta}^2}
  =
  qcC_{T,\mathrm{NC}}^{\dot{\theta}^2}
  =
  \frac{q\pi b^3}{R^2}.
  \label{eq:vawt-dimensional-tangential-force}
\end{gather}

Within the small-angle approximation, the chordwise tangential force acts in the rotor-tangential direction for the present analytical test.
With the present force and torque sign conventions, the total torque per unit span due to the non-circulatory force and pitching moment is
\begin{equation}
  Q_{\mathrm{NC}}
  =
  M_{\mathrm{NC}}
  -
  R T_{\mathrm{NC}}^{\dot{\theta}^2}
  =
  \frac{q\pi b^3}{R}
  -
  \frac{q\pi b^3}{R}
  =
  0.
  \label{eq:vawt-torque-cancellation}
\end{equation}
Thus, the torque due to the tangential force is canceled exactly by the non-circulatory pitching moment.

Combining the circulatory and non-circulatory contributions gives
\begin{equation}
  Q_{\mathrm{tot}}
  =
  Q_{\mathrm{circ}}
  +
  Q_{\mathrm{NC}}
  =
  0.
  \label{eq:vawt-total-torque-cancellation}
\end{equation}
This completes the zero-torque balance for all circulatory and non-circulatory thin-airfoil contributions.

\section{Conclusions and future work}
\label{sec:conclusion}

The present work revisits classical unsteady thin airfoil theory and derives a consistent coefficient formulation for generalized lifting-line methods, including blade-element momentum (BEM), lifting-line (LL), and actuator-line (AL) methods.
These methods combine an outer 3-D rotor and wake problem with an inner 2-D airfoil model at each blade section.
The inner 2-D airfoil model can be viewed as an input--output machine. The outer 3-D flow solution and sectional kinematics provide the relative flow and kinematic histories required by the model, which maps them to the corresponding sectional force and moment.
Starting from the dimensional sectional loads, the derivation traces the circulatory and non-circulatory contributions through conversion to coefficient form, coupling with the 2-D airfoil polar, and projection into the selected lift and drag directions.

The zero-lift-shifted three-quarter-chord onset downwash is used to update the shed wake states, which determine the wake-lagged effective angle of attack and therefore the circulatory force magnitude.
The circulatory force direction and the corresponding decomposition into lift and drag are discussed in detail.
This distinction clarifies the physical meaning of the different flow angles and their roles in the sectional force calculation.
The formulation also clarifies how the classical attached flow contributions are combined with a 2-D airfoil polar without double counting.
The airfoil polar supplies the circulatory lift, profile drag, and pitching moment, while the 2-D shed wake-induced drag, direction-projection contribution, and non-circulatory force and moment contributions are added separately.
The downwash-based deficit-state formulation avoids the separate velocity-rate term required by the angle-state formulation.
Its wake-memory scaling allows the new contribution to the attached-flow lag to be progressively reduced as the flow separates, while the stored state continues to decay.

The coned straight blade calculations show that omitting individual contributions from the complete formulation affects rotor-integrated thrust and power differently.
Building on previous work, the zero-onset VAWT analysis is extended to include all circulatory and non-circulatory thin-airfoil contributions and shows that their total torque contributions cancel exactly in the ideal thin-airfoil limit.

Future work should examine the treatment of the individual aerodynamic contributions under fully separated flow.
In the current implementation, the attached flow circulatory dynamics are disabled when the airfoil is fully separated, while the non-circulatory force and moment contributions are retained.
It should be assessed whether all of these contributions, particularly the torsional rate contribution, remain applicable at very large angles of attack or should be modified or disabled.
This question is relevant to wind turbines in standstill conditions and may improve the modeling of stall-induced vibrations.
Future work should also address a fully consistent treatment of turbulent inflow.




\section*{Data availability}
The 2-D airfoil data used in this article are generated with 2-D fully turbulent RANS computations \citep{IEA37}.






\appendix

\clearpage
\section{Nomenclature}
\label{sec:nomenclature}

\subsection{Roman-letter variables used in the present work}

\begin{table}[!htbp]
\centering
\begin{tabular}{@{}lp{0.76\linewidth}@{}}
\hline
\hline
Symbol & Description\\
\hline
$a$ & dimensionless pitch-axis position relative to the mid-chord point, with the pitch axis at $\xi=ab$ \\
$a^{S}$ & acceleration in sectional coordinate system \\
$A_i$ & amplitude of the $i$-th exponential term in the Wagner-function approximation \\
$A_{\mathrm{proj}}$ & projected rotor area \\
$b$, $c$ & half-chord and chord lengths, with $b=c/2$ \\
$C(k)$, $k$ & Theodorsen function and reduced frequency, with $k=\omega b/U$ \\
$C_N$, $C_T$ & sectional normal- and tangential-force coefficients \\
$C_l$, $C_d$, $C_m$ & sectional lift, drag, and pitching-moment coefficients \\
$C_{l,\alpha}$ & attached-flow lift-curve slope \\
$C_{\mathrm{T,rot}}$, $C_{\mathrm{P,rot}}$ & rotor-integrated thrust and power coefficients \\
$C_{\mathrm{dec},i,n}$ & decay coefficient for indicial state $i$ over time step $n$ \\
$f_{\mathrm{scale}}$, $f_{\mathrm{sep}}$ & scaling factor for the new state increment and lagged dynamic separation-point position \\
$M_a$ & pitching moment per unit span about the pitch axis at $\xi=ab$ \\
$M_{1/4}^{\mathrm{thin}}$ & thin-airfoil pitching moment per unit span about the quarter-chord point at $\xi=-b/2$ \\
$n_w$ & number of exponential terms in the Wagner-function approximation \\
$N$ & normal force per unit span, positive from the pressure side towards the suction side \\
$q$ & dynamic pressure, $q=\rho U^2/2$ \\
$Q$ & aerodynamic torque per unit span about the rotor center \\
$R$ & circular-path radius for a VAWT \\
$R_0$ & baseline rotor radius of a HAWT \\
$s$ & non-dimensional convective time measured in half-chords \\
$t$ & physical time \\
$T$ & tangential force per unit span, positive towards the leading edge \\
$T_0$ & half-chord convection time, $T_0=b/U$ \\
$\mathbf{T}_{G\rightarrow S}$ & transformation matrix from the global coordinate system $G$ to the sectional coordinate system $S$ \\
$T_{\mathrm{rot}}$, $P_{\mathrm{rot}}$ & rotor-integrated thrust and power \\
$U$, $U_0$, $V$ & sectional relative streamwise velocity, onset wind speed, and constant streamwise flow velocity \\
$w_{3/4}$, $w_E$ & three-quarter-chord onset downwash and wake-lagged downwash \\
$\widetilde{w}_{3/4}$, $\widetilde{w}_E$ & corresponding zero-lift-shifted downwash quantities \\
$x$, $y$ & airfoil displacements parallel and normal to the undisturbed flow direction \\
$x_i^w$, $r_i^w$ & direct downwash and downwash-deficit states \\
$z_i$ & angle-of-attack state \\
\hline
\hline
\end{tabular}
\end{table}

\clearpage

\subsection{Greek-letter variables used in the present work}

\begin{table}[!htbp]
\centering
\begin{tabular}{@{}lp{0.76\linewidth}@{}}
\hline
\hline
Symbol & Description\\
\hline
$\alpha_{1/4}$, $\alpha_{3/4}$ & angles of attack at the quarter- and three-quarter-chord points \\
$\alpha_0$ & zero-lift angle of the 2-D airfoil polar \\
$\alpha_E$ & wake-lagged effective angle of attack \\
$\alpha_{\chi}$ & angle of attack at aerodynamic calculation point $\chi$ \\
$\alpha_{\mathrm{circ,dir}}$ & force-direction angle corresponding to zero circulatory drag \\
$\alpha_r$ & angle of the reference direction used to define lift and drag \\
$\beta_i$ & decay coefficient of the $i$-th exponential term in the Wagner-function approximation \\
$\Gamma$ & bound circulation \\
$\chi$ & chordwise location of the aerodynamic calculation point \\
$\Delta s_n$ & increment in non-dimensional semi-chord time over time step $n$ \\
$\Delta t$ & time-step size \\
$\ddot{\epsilon}$ & mid-chord heaving acceleration \\
$\kappa$ & rotor cone angle \\
$\lambda_0$ & baseline rotor tip-speed-ratio, $\lambda_0=\Omega R_0/U_0$ \\
$\phi(s)$ & Wagner indicial-response function \\
$\rho$ & air density \\
$\theta$, $\dot{\theta}$, $\ddot{\theta}$ & pitching angle, effective torsional rate, and torsional acceleration \\
$\xi$ & chordwise coordinate measured from the mid-chord point \\
$\omega$ & angular frequency of the harmonic motion \\
$\dot{\omega}_{x}^{S}, \dot{\omega}_{y}^{S}, \dot{\omega}_{z}^{S}$ & angular-acceleration components about the sectional $x$-, $y$-, and $z$-axes \\
$\Omega$ & rotor rotational speed \\
\hline
\hline
\end{tabular}
\end{table}

\clearpage

\subsection{Subscripts used in the present work}

\begin{table}[!htbp]
\centering
\begin{tabular}{@{}*{2}{l}}
\hline
\hline
Symbol & Description\\
\hline
$1/4$, $3/4$ & quarter- and three-quarter-chord points \\
$\chi$ & aerodynamic calculation point \\
$E$ & equivalent wake-lagged quantity \\
$\mathrm{circ}$ & circulatory contribution \\
$\mathrm{dec}$ & decay coefficient \\
$\mathrm{NC}$ & non-circulatory contribution \\
$\mathrm{ind}$ & 2-D shed wake-induced contribution \\
$\mathrm{proj}$ & projected rotor quantity \\
$r$ & selected reference direction \\
$\mathrm{rot}$ & rotor-integrated quantity \\
$i$ & state index \\
$n$, $n+1$ & discrete time step index \\
\hline
\hline
\end{tabular}
\end{table}

\subsection{Superscripts used in the present work}

\begin{table}[!htbp]
\centering
\begin{tabular}{@{}*{2}{l}}
\hline
\hline
Symbol & Description\\
\hline
$\mathrm{acc}$ & contribution associated with mid-chord heaving acceleration \\
$\mathrm{dyn}$ & unsteady contribution evaluated at the effective angle of attack \\
$\mathrm{full}$ & full expression \\
$G$, $S$ & quantity in the global and sectional coordinate systems, respectively \\
$\mathrm{lin}$ & linear thin-airfoil contribution \\
$\mathrm{polar}$ & quantity obtained directly from the 2-D airfoil polar \\
$\mathrm{profile}$ & profile contribution obtained from the 2-D airfoil polar \\
$\mathrm{suction}$ & leading-edge-suction contribution \\
$\mathrm{thin}$ & quantity obtained from classical thin-airfoil theory \\
$\mathrm{tor}$ & contribution associated with the effective torsional rate \\
$\mathrm{tot}$ & total contribution \\
$\dot{\theta}^2$ & contribution proportional to the square of the effective torsional rate \\
$\Delta\alpha$ & direction-projection contribution \\
\hline
\hline
\end{tabular}
\end{table}

\clearpage

\section{Indicial formulation: downwash and deficit states}
\label{app:indicial-lag-updates}

This appendix explains the zero-lift offset applied to the downwash input and derives the time-discrete updates for the direct downwash-state formulation and the deficit-state formulation with wake-memory scaling.
Following the indicial terminology of \citet{Hansen2004}, each update consists of the decay of the previous state and a new increment.
The deficit-state formulation of \citet{Pirrung2018} uses the downwash as input and applies a scaling factor to the new contribution to the attached-flow lag.
The present work applies the same treatment to the zero-lift-shifted downwash for a cambered airfoil, while the direct downwash-state formulation is included as an alternative attached-flow discretization for reference.

\subsection{Zero-lift offset in the downwash input}
\label{sec:appendix-zero-lift-offset}

\citet{Pirrung2018} write the deficit-state update using the three-quarter-chord onset downwash
\begin{equation}
  w_{3/4}
  =
  U\alpha_{3/4}.
  \label{eq:appendix-geometric-downwash}
\end{equation}
The zero-lift offset associated with airfoil camber is not considered explicitly in that formulation.
Following \citet{Pirrung2017Comparison}, the input to the shed wake model is referenced to the zero-lift angle.
In the present downwash-based formulation, this gives
\begin{equation}
  \widetilde{w}_{3/4}
  =
  U\left(\alpha_{3/4}-\alpha_0\right)
  =
  w_{3/4}-U\alpha_0.
  \label{eq:appendix-shifted-downwash}
\end{equation}

The shed wake memory represents the delayed aerodynamic response associated with vorticity shed from the trailing edge as the bound circulation changes.
For a cambered airfoil in the linear attached-flow range, the circulatory lift per unit span is
\begin{equation}
  L_{\mathrm{circ}}
  =
  \frac{1}{2}
  \rho U^2 c
  C_{l,\alpha}
  \left(
    \alpha_{3/4}-\alpha_0
  \right),
  \label{eq:appendix-attached-circulatory-lift}
\end{equation}
where $C_{l,\alpha}$ is the attached-flow lift-curve slope.
Using the Kutta--Joukowski relation
\begin{equation}
  L_{\mathrm{circ}}
  =
  \rho U\Gamma,
  \label{eq:appendix-kutta-joukowski}
\end{equation}
the corresponding bound circulation is
\begin{equation}
  \Gamma
  =
  \frac{1}{2}
  UcC_{l,\alpha}
  \left(
    \alpha_{3/4}-\alpha_0
  \right).
  \label{eq:appendix-bound-circulation}
\end{equation}
The bound circulation is therefore proportional to the zero-lift-shifted downwash,
\begin{equation}
  \Gamma
  =
  \frac{1}{2}
  cC_{l,\alpha}
  \widetilde{w}_{3/4}.
  \label{eq:appendix-circulation-shifted-downwash}
\end{equation}
For the classical thin-airfoil lift-curve slope $C_{l,\alpha}=2\pi$, this reduces to
\begin{equation}
  \Gamma
  =
  \pi c\widetilde{w}_{3/4}.
  \label{eq:appendix-thin-airfoil-circulation}
\end{equation}
Thus, changes in $\widetilde{w}_{3/4}$ correspond directly to changes in the bound circulation that generate the shed wake memory.

The zero-lift offset also ensures a consistent downwash input when $U$ varies.
Differentiating Eq.~(\ref{eq:appendix-geometric-downwash}) gives
\begin{equation}
  \dot{w}_{3/4}
  =
  U\dot{\alpha}_{3/4}
  +
  \dot{U}\alpha_{3/4}.
  \label{eq:appendix-geometric-downwash-rate}
\end{equation}
Consider a case in which the three-quarter-chord angle of attack remains equal to the constant zero-lift angle $\alpha_0$ while $U$ varies.
The corresponding conditions are
\begin{subequations}
\label{eq:appendix-unshifted-zero-lift-conditions}
\begin{gather}
  \alpha_{3/4}
  =
  \alpha_0,
  \label{eq:appendix-unshifted-zero-lift-angle}
  \\
  \dot{\alpha}_{3/4}
  =
  0.
  \label{eq:appendix-unshifted-zero-lift-angle-rate}
\end{gather}
\end{subequations}
Equation~(\ref{eq:appendix-geometric-downwash-rate}) then gives
\begin{equation}
  \dot{w}_{3/4}
  =
  \dot{U}\alpha_0.
  \label{eq:appendix-unshifted-zero-lift-rate}
\end{equation}
The unshifted onset downwash would therefore produce a new contribution to the attached-flow lag when $U$ varies, even though the attached circulatory lift and bound circulation are zero.

For constant $\alpha_0$, differentiation of Eq.~(\ref{eq:appendix-shifted-downwash}) gives
\begin{equation}
  \dot{\widetilde{w}}_{3/4}
  =
  U\dot{\alpha}_{3/4}
  +
  \dot{U}\left(\alpha_{3/4}-\alpha_0\right).
  \label{eq:appendix-shifted-downwash-rate}
\end{equation}
Under the zero-lift conditions in Eq.~(\ref{eq:appendix-unshifted-zero-lift-conditions}), this gives
\begin{equation}
  \dot{\widetilde{w}}_{3/4}
  =
  0,
  \label{eq:appendix-shifted-zero-lift-rate}
\end{equation}
independently of changes in $U$.

Because $f_{\mathrm{scale}}$ scales the generation of new deficit, the deficit-state update uses $\widetilde{w}_{3/4}$, which is proportional to the bound circulation.
The zero-lift-shifted downwash $\widetilde{w}_{3/4}$ is used in the wake memory update, and the effective angle $\alpha_E$ is recovered before the airfoil polar is evaluated.
\subsection{Non-dimensional semi-chord time and decay coefficient}

The non-dimensional semi-chord time $s$ in Eq.~(\ref{eq:convective-time}) measures the distance travelled by the wake in half-chords.
Differentiation with respect to physical time gives
\begin{equation}
  \frac{\mathrm ds}{\mathrm dt}
  =
  \frac{2U(t)}{c}
  =
  \frac{1}{T_0(t)}.
  \label{eq:appendix-convective-time-rate}
\end{equation}
Integrating Eq.~(\ref{eq:appendix-convective-time-rate}) from $t_n$ to $t_{n+1}$ gives the exact increment in non-dimensional time
\begin{equation}
  \Delta s_n
  =
  s_{n+1}-s_n
  =
  \int_{t_n}^{t_{n+1}}\frac{\mathrm dt}{T_0(t)}
  =
  \frac{2}{c}\int_{t_n}^{t_{n+1}}U(t)\,\mathrm dt.
  \label{eq:appendix-ds}
\end{equation}
In the HAWC2 implementation, this increment is approximated using the relative velocity at the end of the time step \citep{Pirrung2018}:
\begin{equation}
  \Delta s_n
  =
  \frac{2U_{n+1}\Delta t_n}{c}.
  \label{eq:appendix-ds-endpoint}
\end{equation}

The decay coefficient follows from the homogeneous deficit-state equation in non-dimensional time:
\begin{equation}
  \frac{\mathrm d r_i^w}{\mathrm ds}
  +
  \beta_i r_i^w
  =
  0.
  \label{eq:appendix-homogeneous-deficit-convective}
\end{equation}
Its exact solution over one time step is
\begin{subequations}
\label{eq:appendix-deficit-decay}
\begin{gather}
  r_{i,n+1}^w
  =
  C_{\mathrm{dec},i,n}r_{i,n}^w,
  \label{eq:appendix-homogeneous-deficit-update}
  \\
  C_{\mathrm{dec},i,n}
  =
  \exp\left(-\beta_i\Delta s_n\right).
  \label{eq:appendix-direct-decay-factor}
\end{gather}
\end{subequations}
For $\beta_i>0$ and $\Delta s_n\geq0$, the decay coefficient $C_{\mathrm{dec},i,n}$ is the fraction of the stored state retained during the time step.

\subsection{Direct downwash-state update}

Using $\mathrm ds/\mathrm dt=T_0^{-1}$, the direct state equation for the zero-lift-shifted downwash becomes
\begin{equation}
  \frac{\mathrm d x_i^w}{\mathrm ds}
  +
  \beta_i x_i^w
  =
  \beta_iA_i\widetilde{w}_{3/4}.
  \label{eq:appendix-direct-convective-ode}
\end{equation}
Here and below, $i=1,\ldots,n_w$.
Multiplication by $e^{\beta_i s}$ and integration over one time step give the indicial update
\begin{equation}
  x_{i,n+1}^w
  =
  C_{\mathrm{dec},i,n}x_{i,n}^w
  +
  \beta_iA_i
  \int_{s_n}^{s_{n+1}}
  e^{-\beta_i\left(s_{n+1}-\sigma\right)}
  \widetilde{w}_{3/4}(\sigma)\,\mathrm d\sigma.
  \label{eq:appendix-direct-integral-solution}
\end{equation}
Following the terminology of \citet{Hansen2004}, the first term is the decayed previous state and the integral term is the new increment.
Assuming that $\widetilde{w}_{3/4}$ is piecewise constant over the time step and evaluating it at the end of the step give
\begin{equation}
  x_{i,n+1}^w
  =
  C_{\mathrm{dec},i,n}x_{i,n}^w
  +
  A_i\left(1-C_{\mathrm{dec},i,n}\right)
  \widetilde{w}_{3/4,n+1}.
  \label{eq:appendix-direct-state-update}
\end{equation}
The zero-lift-shifted equivalent wake-lagged downwash and the corresponding effective angle of attack are then
\begin{gather}
  \widetilde{w}_{E,n+1}
  =
  \left(1-\sum_{i=1}^{n_w}A_i\right)\widetilde{w}_{3/4,n+1}
  +
  \sum_{i=1}^{n_w}x_{i,n+1}^w,
  \label{eq:appendix-direct-wE}
  \\
  \alpha_{E,n+1}
  =
  \alpha_0
  +
  \frac{\widetilde{w}_{E,n+1}}{U_{n+1}}.
  \label{eq:appendix-direct-alphaE}
\end{gather}

\subsection{Deficit-state update with wake-memory scaling}

The deficit-state formulation introduced in Sect.~\ref{sec:wake-state} describes the attached-flow wake-memory dynamics according to Eq.~(\ref{eq:w-deficit-ode}).
For the zero-lift-shifted downwash, the deficit state is defined as
\begin{equation}
  r_i^w
  =
  A_i\widetilde{w}_{3/4}
  -
  x_i^w.
  \label{eq:appendix-deficit-transform-shifted}
\end{equation}
Using $\mathrm ds/\mathrm dt=T_0^{-1}$, Eq.~(\ref{eq:w-deficit-ode}) can be written in non-dimensional time as
\begin{equation}
  \frac{\mathrm d r_i^w}{\mathrm ds}
  +
  \beta_i r_i^w
  =
  A_i
  \frac{\mathrm d\widetilde{w}_{3/4}}{\mathrm ds}.
  \label{eq:appendix-unscaled-deficit-convective-ode}
\end{equation}

When this formulation is embedded in the full dynamic-stall model, the separation state scales the generation of new attached-flow wake memory as the flow separates~\citep{Pirrung2018}.
Specifically, the scaling function $f_{\mathrm{scale}}$ multiplies the new contribution to the attached-flow lag:
\begin{equation}
  \dot{r}_i^w
  +
  \frac{\beta_i}{T_0}r_i^w
  =
  A_i f_{\mathrm{scale}}
  \dot{\widetilde{w}}_{3/4}.
  \label{eq:appendix-w-deficit-ode-scaled}
\end{equation}
The term $\beta_i r_i^w/T_0$ describes the decay of the stored state, whereas
$A_i f_{\mathrm{scale}}\dot{\widetilde{w}}_{3/4}$ describes the scaled generation of new attached-flow wake memory.

In the discrete HAWC2 dynamic-stall coupling, the scaling factor used to calculate the aerodynamic states at time $t$ is evaluated using the converged dynamic separation-point position from the previous time step:
\begin{equation}
  f_{\mathrm{scale}}(t)
  =
  f_{\mathrm{sep}}(t-\Delta t).
  \label{eq:appendix-fscale}
\end{equation}
Here, $f_{\mathrm{sep}}=1$ represents fully attached flow, whereas $f_{\mathrm{sep}}=0$ represents fully separated flow.

For fully attached flow, $f_{\mathrm{scale}}=1$, and Eq.~(\ref{eq:appendix-w-deficit-ode-scaled}) reduces to the attached-flow deficit-state equation, Eq.~(\ref{eq:w-deficit-ode}).
As separation develops and $f_{\mathrm{scale}}$ decreases, the generation of new wake memory is progressively reduced, while the stored state continues to decay with its original time scale.
For fully separated flow, $f_{\mathrm{scale}}=0$, so no new wake-memory contribution is generated.

Using $\mathrm ds/\mathrm dt=T_0^{-1}$, Eq.~(\ref{eq:appendix-w-deficit-ode-scaled}) becomes
\begin{equation}
  \frac{\mathrm d r_i^w}{\mathrm ds}
  +
  \beta_i r_i^w
  =
  A_i f_{\mathrm{scale}}
  \frac{\mathrm d\widetilde{w}_{3/4}}{\mathrm ds}.
  \label{eq:appendix-deficit-convective-ode}
\end{equation}

Assuming that $\widetilde{w}_{3/4}$ varies linearly in non-dimensional time over the step and that $f_{\mathrm{scale},n}$ remains constant, integration of Eq.~(\ref{eq:appendix-deficit-convective-ode}) gives the indicial update
\begin{equation}
  r_{i,n+1}^w
  =
  C_{\mathrm{dec},i,n}r_{i,n}^w
  +
  f_{\mathrm{scale},n}
  \frac{A_i}{\beta_i}
  \frac{1-C_{\mathrm{dec},i,n}}{\Delta s_n}
  \left(
    \widetilde{w}_{3/4,n+1}
    -
    \widetilde{w}_{3/4,n}
  \right).
  \label{eq:appendix-deficit-step}
\end{equation}
The first term is the decayed previous state, and the second term is the scaled new increment.

Using the end-of-step approximation for the non-dimensional time increment in Eq.~(\ref{eq:appendix-ds-endpoint}), the update becomes
\begin{equation}
  r_{i,n+1}^w
  =
  C_{\mathrm{dec},i,n}r_{i,n}^w
  +
  f_{\mathrm{scale},n}
  \frac{A_i}{\beta_i}
  \frac{T_{0,n+1}}{\Delta t_n}
  \left(
    1-C_{\mathrm{dec},i,n}
  \right)
  \left(
    \widetilde{w}_{3/4,n+1}
    -
    \widetilde{w}_{3/4,n}
  \right).
  \label{eq:appendix-deficit-physical-time-step}
\end{equation}

The zero-lift-shifted equivalent wake-lagged downwash and the corresponding effective angle of attack are then
\begin{gather}
  \widetilde{w}_{E,n+1}
  =
  \widetilde{w}_{3/4,n+1}
  -
  \sum_{i=1}^{n_w}r_{i,n+1}^w,
  \label{eq:appendix-deficit-wE}
  \\
  \alpha_{E,n+1}
  =
  \alpha_0
  +
  \frac{\widetilde{w}_{E,n+1}}{U_{n+1}}
  =
  \alpha_{3/4,n+1}
  -
  \frac{1}{U_{n+1}}
  \sum_{i=1}^{n_w}r_{i,n+1}^w.
  \label{eq:appendix-deficit-alphaE}
\end{gather}
The zero-lift offset therefore modifies the downwash input used to advance the wake-lag states, while $\alpha_E$ remains the effective angle supplied to the airfoil polar.









\section*{Author contributions}
This work builds on previous research by MG, GRP, and AL on the incompressible Beddoes--Leishman type dynamic stall model and its implementation in HAWC2.
The complete attached flow formulation was described by AL, with contributions from MG and GRP.
The interpretation of classical unsteady thin-airfoil theory and its coupling with airfoil polars for wind turbine applications were primarily developed by MG, with contributions from GRP and AL.
The deficit-state formulation and its HAWC2 implementation were primarily developed by GRP and MG, with contributions from AL.
The zero-onset VAWT case was derived by AL, with contributions from MG and GRP.
AL performed the numerical studies using CFD, lifting-line, and BEM models.
AL post-processed and discussed the numerical results, with contributions from MG and GRP.
All authors contributed to the conclusions.
Future work directions were primarily proposed by GRP, with contributions from MG and AL.
AL wrote the original draft. All authors contributed to reviewing and revising the manuscript and approved the final manuscript.

\section*{Competing interests}
DTU Wind and Energy Systems develops and distributes the Navier--Stokes solver EllipSys3D on commercial and academic terms.
DTU Wind and Energy Systems also develops, supports and distributes HAWC2 on commercial terms, and HAWC2 is available free of charge for educational and academic research purposes.


\section*{Acknowledgements}
  The authors gratefully acknowledge David Marten of QBlade.org Consulting for valuable discussions on this topic.
  The authors would like to thank Frederik Zahle of DTU Wind and Energy Systems for his contributions to establishing the fully-scripted mesh generation and post-processing workflows for the Navier--Stokes simulations in EllipSys3D.
  Computational and data resources were provided by the Sophia HPC Cluster at DTU (https://doi.org/10.57940/fafc-6m81, DTU, 2019).
%
  Earlier versions of this paper were revised iteratively with the assistance of AI tools, including OpenAI's ChatGPT (GPT-5.5 and GPT-5.6). These tools were used by the authors to generate suggestions for improving language and wording based on author-prepared drafts that were developed and updated throughout the preparation of the paper.
  The paper underwent multiple rounds of revision by the authors, with AI-generated suggestions being selectively incorporated and extensively modified.

\bibliographystyle{abbrvnat}
\bibliography{references}
\end{document}